\documentclass[11pt]{article}
\usepackage{epsfig}
\usepackage{amsmath}
\usepackage{array}
\usepackage{rotating}
\usepackage{float}
\usepackage{multirow}
\usepackage{graphicx}

\usepackage[hidelinks]{hyperref} 

\makeatletter
\def\singlespace{\def\baselinestretch{1}\@normalsize}

\@addtoreset{equation}{section}
\renewcommand{\baselinestretch}{1.412}

\renewcommand{\theequation}{\arabic{section}.\arabic{equation}}

\newcommand{\wh}{\widehat}
\newcommand{\wt}{\widetilde}

\newcommand{\diag}{\mbox{diag}}

\newcommand{\T}{\!\mbox{\scriptsize T}}

\newcommand{\bbE}{E}
\newcommand{\cov}{\mbox{cov}}

\newcommand{\SR}{\mbox{SR}}

\newcommand{\CV}{\mbox{CV}}
\newcommand{\SD}{\mbox{SD}}
\newcommand{\GARCH}{\mbox{GARCH}}

\newcommand{\calF}{{\cal F}}
\newcommand{\calW}{{\cal W}}
\newcommand{\calX}{{\cal X}}

\newcommand{\calQ}{{\cal Q}}

\newcommand{\bX}{X}

\newcommand{\ep}{\mbox{\boldmath{$\epsilon$}}}
\newcommand{\bbeta}{\mbox{\boldmath{$\beta$}}}
\newcommand{\bxi}{\mbox{\boldmath{$\xi$}}}
\newcommand{\bphi}{\mbox{\boldmath{$\Phi$}}}
\newcommand{\bPhi}{\mbox{\boldmath{$\Phi$}}}
\newcommand{\btheta}{\mbox{\boldmath{$\theta$}}}
\newcommand{\bGamma}{\mbox{\boldmath{$\Gamma$}}}
\newcommand{\bLambda}{\mbox{\boldmath{$\Lambda$}}}
\newcommand{\bTheta}{\mbox{\boldmath{$\Theta$}}}

\newcommand{\bmu}{\mbox{\boldmath{$\mu$}}}
\newcommand{\bsigma}{\mbox{\boldmath{$\sigma$}}}
\newcommand{\tbX}{\widetilde{X}}
\newcommand{\bu}{\mbox{\bf u}}
\newcommand{\bbP}{P}
\newcommand{\etal}{{\it et al.}\/}
\newcommand{\bY}{Y}

\newcommand{\by}{\mbox{\bf y}}
\newcommand{\bx}{\mbox{\bf x}}
\newcommand{\ba}{\mbox{\bf a}}
\newcommand{\bB}{\mbox{\bf B}}

\newcommand{\bA}{\mbox{\bf A}}

\newcommand{\bg}{\mbox{\bf g}}

\newcommand{\bc}{\mbox{\bf c}}
\newcommand{\bF}{\mbox{\bf F}}
\newcommand{\tK}{\widetilde{K}}
\newcommand{\br}{\mbox{\bf r}}
\newcommand{\bbX}{\mbox{\bf X}}
\newcommand{\bV}{\mbox{\bf V}}
\newcommand{\bU}{\mbox{\bf U}}
\newcommand{\bb}{\mbox{\bf b}}
\newcommand{\bbb}{\textbf{b}}
\newcommand{\be}{\mbox{\bf e}}
\newcommand{\bR}{\mbox{\bf R}}
\newcommand{\bw}{\mbox{\bf w}}
\newcommand{\0}{\mbox{\bf 0}}
\newcommand{\1}{\mbox{\bf 1}}

\newcommand{\tr}{\mbox{tr}}

\newcommand{\bSigma}{\boldsymbol{\Sigma}}
\newcommand{\bE}{\mathbf{E}}

\newcommand{\bbbeta}{\boldsymbol{\beta}}
\newcommand{\bbtheta}{\boldsymbol{\theta}}
\newcommand{\bbTheta}{\boldsymbol{\Theta}}
\newcommand{\bbLambda}{\boldsymbol{\Lambda}}

\LetLtxMacro{\@stdmakecaption}{\@makecaption}
\renewcommand{\@makecaption}[2]{\bfseries\@stdmakecaption{#1}{#2}}

\begin{document}
\title{\LARGE \bf
A Dynamic Structure for High Dimensional Covariance Matrices and its
Application in Portfolio Allocation
\footnote{This research is supported by the Singapore National Research
Foundation under its Cooperative Basic Research Grant and administered by the
Singapore Ministry of Health's National Medical Research Council (Grant No.
NMRC/CBRG/0014/2012) and the National Natural Science Foundation of China
(Grant No. 11271242). John Box was supported by an EPSRC funded studentship
through the University of York. Shaojun Guo was partly supported by Key
Laboratory of RCSDS, Chinese Academy of Sciences and an EPSRC research grant in United Kingdom.}
      }
\author{
Shaojun Guo
\\
Chinese Academy of Sciences,
Beijing, People's Republic of China
\\
\& London School of Economics,
London, United Kingdom
\and
John Box
\\
Department of Mathematics
\\
University of York, United Kingdom
\and
Wenyang Zhang
\\
Department of Mathematics
\\
University of York, United Kingdom
}
\maketitle

\begin{abstract}

Estimation of high dimensional covariance matrices is an interesting and
important research topic.  In this paper, we propose a dynamic structure 
and develop an estimation procedure for high 
dimensional covariance matrices.  Asymptotic
properties are derived to justify the estimation procedure and simulation 
studies are conducted to demonstrate its performance when the sample size is 
finite.  By exploring a financial application, an
empirical study shows that portfolio allocation based on dynamic high 
dimensional covariance matrices can significantly outperform the market 
from 1995 to 2014.  Our proposed method also outperforms portfolio allocation 
based on the sample covariance matrix and the portfolio allocation 
proposed in Fan, Fan and Lv (2008).

\vspace{0.2in}


{\small \bf SHORT TITLE}:  Dynamic Structure for HDCM.

\end{abstract}

\section{Introduction}
\label{intro}

Covariance matrix estimation is an important topic 
in statistics and econometrics with wide applications 
in many disciplines, such as economics, 
finance and psychology.  A traditional approach
 to estimating covariance matrices is based on the sample 
covariance matrix.  However, the sample covariance matrix would 
not be a good choice when the dimension is large, and 
especially when the inverse is required, which is often the 
case when constructing a portfolio allocation 
in finance.  This is because the
estimation errors would accumulate when using the inverse of the sample
covariance matrix to estimate the inverse of the covariance matrix.  When the
size of the covariance matrix is large, the cumulative estimation error would
become unacceptable even if the estimation error of each entry of the
covariance matrix is tiny.

In recent years there has been various attempts to address high dimensional 
covariance matrix estimation.  Usually, a sparsity condition is imposed 
to control the trade-off between variance and bias.  See, Wu and
Pourahmadi (2003), El Karoui (2008), Bickel and Levina (2008a, 2008b), Lam and
Fan (2009), Fan, Liao, and Mincheva (2011), and the references 
therein.  Fan, Fan and Lv (2008) considered a different approach by imposing 
a factor model and estimated the covariance matrix based on this structure.

Most of the literature addressing high dimensional covariance matrix 
estimation assumes that the covariance matrix is constant over time.  However, in many
applications, covariance matrices are dynamic.  For example, 
today's optimal portfolio allocation may not be 
optimal tomorrow, or next month.  Therefore, when applying the formula for 
Markowitz's optimal portfolio allocation (Markowitz 1959), the covariance matrix used should 
be dynamic and allowed to change 
over time.

In order to introduce a dynamic structure for covariance matrices, 
one cannot simply assume each entry of a covariance matrix is a 
function of time because this would not serve very well in 
prediction.  Instead, we start with an approach stimulated 
by Fan, Fan and Lv (2008) which is based on the Fama-French three-factor 
model (Fama and French, 1992, 1993)
\begin{equation}
y_t = \alpha + X_t^{\T} \ba + \epsilon_t,
\label{ff}
\end{equation}
where $y_t$ is the excess return of an asset and $X_t$ is 
the vector of the three factors at time $t$. 
To make (\ref{ff}) more flexible, we allow $\ba$ to depend on the values 
of the three factors at time $t-1$.  To avoid the so-called 
`curse of dimensionality', we assume this dependence is 
through a linear combination of the values of the three 
factors at time $t-1$, which brings us to
\begin{equation}
y_t
=
\alpha(X_{t-1}^{\T} \bbeta)  + X_t^{\T} \ba(X_{t-1}^{\T} \bbeta) + \epsilon_t.
\label{sff}
\end{equation}
This motivates a dynamic structure for the covariance matrix 
of a random vector $Y_t$ through an adaptive varying 
coefficient model which we shall now introduce.

Suppose $(X_t^{\T}, \ Y_t^{\T})$, $t=1, \ \cdots, \ n$, is a time series, 
where $Y_t$ is a $p_n$ dimensional vector and $X_t$ is a $q$ dimensional 
factor. An underlying assumption throughout this paper is that 
$p_n \longrightarrow \infty$ when $n \longrightarrow \infty$, and $q$ is 
fixed. Also, we assume that $X_t$, $t=1, \ \cdots, \ n$, is a stationary Markov process.
We assume
\begin{equation}
Y_t = \bg(X_{t-1}^{\T} \bbeta) + \bphi(X_{t-1}^{\T} \bbeta) X_t + \ep_t,
\quad
\|\bbeta \| = 1,
\quad
\beta_1 > 0
\label{model}
\end{equation}
where $\bbeta = (\beta_1, \ \cdots, \ \beta_q)^{\T}$, $\bphi(X_{t-1}^{\T} \bbeta)$ is a factor loading matrix which is varying over $X_{t-1}^{\T} \bbeta$, and $\{\ep_t$, $t=1, \ \cdots, \ n\}$ are random errors which are independent of $\{X_t$, $t=1, \ \cdots, \ n\}$.  We assume
$$
E(\ep_t|\{\ep_l: \ l < t\}) = \0,
\quad
\cov(\ep_t|\{\ep_l: \ l < t\})
=
\bSigma_{0,t}
=
\diag\left(
\sigma_{1t}^2, \ \cdots, \ \sigma_{p_n t}^2
\right)
$$
where
\begin{equation}
\sigma_{k t}^2
=
\alpha_{k,0} + \sum\limits_{i=1}^m \alpha_{k,i} \epsilon_{k,t-i}^2
+ \sum\limits_{j=1}^s \gamma_{k,j} \sigma_{k,t-j}^2,
\quad
t = 2, \ \cdots, \ n,
\label{gar}
\end{equation}
for each $k=1, \ \cdots, \ p_n$ and for some integers $m$ and $s$.
Let $\calF_t$ be the $\sigma-$algebra generated 
by  $\{(X_l^{\T}, \ \ep_l^{\T}): \ l \leq t\}$. 
The main focus of this paper is on the conditional covariance matrix
\begin{equation}
\cov(Y_t|\calF_{t-1})
=
\bphi(X_{t-1}^{\T} \bbeta) \bSigma_x(X_{t-1}) \bphi(X_{t-1}^{\T} \bbeta)^{\T}
+
\bSigma_{0,t}
\label{goal}
\end{equation}
where $\bSigma_x(X_{t-1}) = \cov(X_t|X_{t-1}).$ 
In (\ref{goal}), $\bphi(\cdot)$, $\bbeta$, $\bSigma_x(\cdot)$, $\alpha_{k,i}$ and
$\gamma_{k,j}$, $i=0, \ \cdots, \ m$, $j=1, \ \cdots, \ s$, are unknown and need to be
estimated.  Not only does (\ref{goal}) introduce a dynamic structure for
$\cov(Y_t|\calF_{t-1})$, but also reduces the number of unknown parameters
from $p_n(p_n+1)/2$ to $p_n q + q^2$ unknown functions and $q + s + m + 1$
unknown parameters.

We remark that model (\ref{model}) is interesting in its own 
right, since it combines single-index modelling (Carroll \etal, 1997, 
H\"{a}rdle \etal, 1993, Yu and Ruppert, 2002, 
Xia and H\"{a}rdle, 2006, Kong and Xia, 2014) and 
varying coefficient modelling (Fan and Zhang, 1999, 2000, 
Fan \etal, 2003,
Sun \etal, 2007, Zhang \etal, 2009, Li and Zhang, 2011, 
Sun \etal, 2014).  In this paper, as a by-product,
an estimation procedure for (\ref{model}) is proposed and an iterative algorithm
is developed for implementation purposes.

This paper is organised as follows.  We begin in Section \ref{est0} with a
description of the proposed estimation procedure 
for $\cov(Y_t|\calF_{t-1})$.  A discussion on bandwidth 
selection is given in Section \ref{BS}.  In
Section \ref{asy} we provide asymptotic properties of the
estimation procedure.  An iterative algorithm to implement the
estimation procedure is suggested in Section \ref{alg}.  Using 
the proposed dynamic structure for covariance matrices and the 
developed estimation procedure, we outline a process for constructing a 
portfolio allocation based on the formula for Markowitz's 
optimal portfolio in Section \ref{pa}.  The performance 
of the estimation procedure and portfolio allocation 
are also assessed by simulation studies in
Section \ref{sim}.  In Section \ref{real}, we apply the portfolio allocation 
methodology to a data set consisting of 49 industry 
portfolios which are freely available from Kenneth 
French's website. We find that the proposed methodology 
works surprisingly well. All the detailed proofs are relegated to the appendix.

\section{Estimation procedure}
\label{est0}

In this section, we are going to introduce an estimation procedure for
$\cov(Y_t|\calF_{t-1})$.  We will first estimate $\bbeta$, $\bphi(\cdot)$,
$\bSigma_x(\cdot)$, $\alpha_{k,i}$ and $\gamma_{k,j}$, and denote the resulting estimators
by $\hat{\bbeta}$, $\hat{\bphi}(\cdot)$, $\hat{\bSigma}_x(\cdot)$,
$\hat{\alpha}_{k,i}$ and $\hat{\gamma}_{k,j}$ for $i=0, \ \cdots, \ m$ and
$j=1, \ \cdots, \ s$.  Let $\hat{\bSigma}_{0,t}$ be $\bSigma_{0,t}$ with
$\alpha_{k,i}$ and $\gamma_{k,j}$ being replaced by $\hat{\alpha}_{k,i}$ and
$\hat{\gamma}_{k,j}$ respectively.  We use
\begin{equation}
\wh{\cov}(Y_t|\calF_{t-1})
=
\hat{\bphi}(X_{t-1}^{\T} \hat{\bbeta})
\hat{\bSigma}_x(X_{t-1})
\hat{\bphi}(X_{t-1}^{\T} \hat{\bbeta})^{\T} +
\hat{\bSigma}_{0,t}
\label{est}
\end{equation}
to estimate $\cov(Y_t|\calF_{t-1})$.

Throughout this paper, for any function $f(x)$, we use $\dot{f}(x)$ to
denote its derivative.  For any functional matrix $F = (f_{ij}(x))$, we define
its derivative as $\dot{F} = (\dot{f}_{ij}(x))$.  For any integers $p$ and
$q$, we use $\0_{p \times q}$ to denote a $p \times q$ matrix with each entry
being $0$, and $\1_p$ to denote a $p$-dimensional vector with each component being $1$.

\subsection{Estimation of $\bbeta$}

A Taylor expansion gives, for $X_i^{\T} \bbeta$ in a neighbourhood of $X_j^{\T} \bbeta$,
$$
\bphi(X_i^{\T} \bbeta)
\approx
\bphi(X_j^{\T} \bbeta) + \dot{\bphi}(X_j^{\T} \bbeta) (X_i - X_j)^{\T} \bbeta
$$
and
$$
\bg(X_i^{\T} \bbeta)
\approx
\bg(X_j^{\T} \bbeta) + \dot{\bg}(X_j^{\T} \bbeta) (X_i - X_j)^{\T} \bbeta
$$
for $j=1, \ \cdots, \ n-1$.  This,
together with the idea of least squares estimation, brings us to the following local
discrepancy function
\begin{eqnarray}
&  &
L(\bg_1, \ \bxi_1, \ A_1, \ B_1, \ \cdots, \ \bg_{n-1}, \ \bxi_{n-1}, \
A_{n-1}, \ B_{n-1}, \ \bbeta)
\nonumber
\\
& = &
\sum\limits_{j=1}^{n-1}
\sum\limits_{i=2}^n
\left\|
Y_i - \bg_j - A_j X_i - (\bxi_j + B_j X_i)(X_{i-1} - X_j)^{\T} \bbeta
\right\|^2 K_h((X_{i-1}- X_j)^{\T} \bbeta),
\label{ls}
\end{eqnarray}
where: $K_h(\cdot) = K(\cdot/h)/h$, $K(\cdot)$ is a kernel function;
$h$ is a bandwidth; and  $\bg_j$, $\bxi_j$, $A_j$ and $B_j$ are used to denote
$\bg(X_j^{\T} \bbeta)$, $\dot{\bg}(X_j^{\T} \bbeta)$, $\bphi(X_j^{\T} \bbeta)$
and $\dot{\bphi}(X_j^{\T} \bbeta)$ respectively.  By minimising
$$
L(\bg_1, \ \bxi_1, \ A_1, \ B_1, \ \cdots, \ \bg_{n-1}, \ \bxi_{n-1}, \
A_{n-1}, \ B_{n-1}, \ \bbeta)
$$
under the conditions
$$
\|\bbeta \| = 1,
\quad
\beta_1 > 0,
$$
we use the corresponding value of $\bbeta$ as the estimator and denote it by $\hat{\bbeta}$.

\subsection{Estimation of $\bphi(\cdot)$ and $\bg(\cdot)$}

Once an estimate $\hat{\bbeta}$ has been obtained, the estimators of $\bphi(\cdot)$
and $\bg(\cdot)$ can be constructed row by row through a standard univariate
varying coefficient model for each component of $Y_t$.  Let
$$
\bg(\cdot)
=
\left(g_1(\cdot), \ \cdots, \ g_{p_n}(\cdot) \right)^{\T},
\quad
\bphi(\cdot)
=
\left(\ba_1(\cdot), \ \cdots, \ \ba_{p_n}(\cdot) \right)^{\T},
\quad
Y_t = (y_{1,t}, \ \cdots, \ y_{p_n, t})^{\T}.
$$
By (\ref{model}), and for $k = 1, \ \cdots, \ p_n$, we have the following synthetic
univariate varying coefficient model
$$
y_{k,t}
=
g_k(X_{t-1}^{\T} \hat{\bbeta}) + X_t^{\T} \ba_k(X_{t-1}^{\T} \hat{\bbeta})
+ \epsilon_{kt},
$$
for $t = 2, \ \cdots, \ n.$
By local linear estimation for standard varying-coefficient models, and for
any given $u$, we have
$$
\hat{\ba}_k(u)
=
(I_q, \ \0_{q\times (q+2)})
\left(\calX^{\T} W \calX \right)^{-1} \calX^{\T} W \by_k,
\quad
\hat{g}_k(u)
=
(\0_{1\times q}, \ 1, \ \0_{1 \times (q+1)})
\left(\calX^{\T} W \calX \right)^{-1} \calX^{\T} W \by_k,
$$
where
$$
\by_k
=
(y_{k,2}, \ \cdots, \ y_{k,n})^{\T},
\quad
\calX
=
\left(
\begin{array}{cccc}
X_2^{\T} & 1 & (X_1^{\T} \hat{\bbeta} - u) &
(X_1^{\T} \hat{\bbeta} - u) X_2^{\T}
\\
\vdots & \vdots & \vdots & \vdots
\\
X_n^{\T} & 1 & (X_{n-1}^{\T} \hat{\bbeta} - u) &
(X_{n-1}^{\T} \hat{\bbeta} - u) X_n^{\T}
\end{array}
\right),
$$
$$
W
=
\diag\left(
K_{h_1}(X_1^{\T} \hat{\bbeta} - u),
\ \cdots, \
K_{h_1}(X_{n-1}^{\T} \hat{\bbeta} - u)
\right),
$$
and $h_1$ is a bandwidth.

\subsection{Estimation of $\bSigma_x(\cdot)$}

In order to estimate
$E(X_t|X_{t-1} = \bu)$ and $E(X_t X_t^{\T}|X_{t-1} = \bu)$, for 
any given $\bu$, we use the local constant estimators
\begin{equation}
\wh{E}(X_t|X_{t-1} = \bu)
=
\frac{\sum\limits_{t=2}^n X_t K_{h_2}(\|X_{t-1} - \bu\|)}
{\sum\limits_{t=2}^n K_{h_2}(\|X_{t-1} - \bu\|)},
\label{mean}
\end{equation}
$$
\wh{E}(X_t X_t^{\T}|X_{t-1} = \bu)
=
\frac{\sum\limits_{t=2}^n X_t X_t^{\T} K_{h_2}(\|X_{t-1} - \bu\|)}
{\sum\limits_{t=2}^n K_{h_2}(\|X_{t-1} - \bu\|)}.
$$
This gives us the following estimator of $\bSigma_x(\bu)$
\begin{eqnarray}
\hat{\bSigma}_x(\bu)
& = &
\wh{E}(X_t X_t^{\T}|X_{t-1} = \bu)
-
\wh{E}(X_t|X_{t-1} = \bu)
\left\{
\wh{E}(X_t|X_{t-1} = \bu)
\right\}^{\T}
\nonumber
\\
& = &
\left\{\tr(\calW)\right\}^{-2}
\bbX^{\T}
\left\{
\tr(\calW) \calW  - \calW \1 \1^{\T} \calW
\right\}
\bbX
\label{sigx}
\end{eqnarray}
where
$$
\bbX
=
(X_2, \ \cdots, \ X_n)^{\T},
\quad
\calW
=
\diag(K_{h_2}(\|X_1 - \bu\|), \ \cdots, \ K_{h_2}(\|X_{n-1} - \bu\|)),
$$
and $h_2$ is a bandwidth.

\subsection{Estimation of $\bSigma_{0,t}$}

For each $k$, $k=1, \ \cdots, \ p_n$, let
$$
r_{k,t}
=
\hat{\epsilon}_{k,t}
=
y_{k,t} - \hat{g}_k(X_{t-1}^{\T} \hat{\bbeta})-
X_t^{\T} \hat{\ba}_k(X_{t-1}^{\T} \hat{\bbeta}).
$$
By (\ref{gar}), we have the following synthetic GARCH model
\begin{equation}
\sigma_{k t}^2
=
\alpha_{k,0} + \sum\limits_{i=1}^m \alpha_{k,i} r_{k,t-i}^2
+ \sum\limits_{j=1}^s \gamma_{k,j} \sigma_{k,t-j}^2,
\quad
t = 2, \ \cdots, \ n
\label{rec}
\end{equation}
which is equivalent to
$$
r_{k,t}^2
=
\alpha_{k,0} +
\sum\limits_{i=1}^{\max(m,s)} (\alpha_{k,i} + \gamma_{k,i}) r_{k,t-i}^2
+ \eta_{kt} - \sum\limits_{j=1}^s \gamma_{k,j} \eta_{k,t-j},
\quad
t = 2, \ \cdots, \ n
$$
where $\eta_{kt} = r_{k,t}^2 - \sigma_{k t}^2$, $\gamma_{k,i} = 0$ when $i>s$,
and $\alpha_{k,i} = 0$ when $i>m$.

Once $\alpha_{k,i}$ and $\gamma_{k,j}$ have been estimated, by substituting
them into (\ref{rec}) and setting
$\sigma_{k l}^2 = r_{k,l}^2$ for $l \leq \max(m,s)$, we can obtain an estimator
$\hat{\sigma}_{k t}^2$ of $\sigma_{k t}^2$ and hence an estimator
$\hat{\bSigma}_{0,t}$ of $\bSigma_{0,t}$.

For each $k$, $k = 1, \ \cdots, \ p_n$, let
$\btheta_k
=
(\alpha_{k,0}, \ \cdots, \ \alpha_{k,m}, \ \gamma_{k,1}, \ \cdots, \
\gamma_{k,s})^{\T}$.  We are going to use a quasi-maximum likelihood approach
to estimate $\btheta_k$.  We define the negative quasi log-likelihood function
of $\btheta_{k}$ as
\begin{equation}
Q_{k,n}(\btheta_k)
=
{n^{-1}} \sum_{t = 2}^n
\left\{
\frac{r_{k,t}^2}{\sigma_{k,t}^2(\btheta_k)}
+
\log \sigma_{k,t}^2(\btheta_k)
\right\}
\label{qmle}
\end{equation}
where $\sigma_{k,t}^2(\btheta_k)$ are recursively defined by (\ref{rec}) with
initial values being either
$$
r_{k,0}^2
=
\cdots
=
r_{k,1-m}^2
=
\sigma_{k,0}^2
=
\cdots
=
\sigma_{k,1-s}^2 = \alpha_{k,0}
$$
or
$$
r_{k,0}^2
=
\cdots
=
r_{k,1-m}^2
=
\sigma_{k,0}^2
=
\cdots
=
\sigma_{k,1-s}^2
=
r_{k,0}^2.
$$
By minimising $Q_{k,n}(\btheta_k)$ with respect to $\btheta_k$ on a
compact set $\bLambda$ defined in (B3) in Appendix A, we use the minimiser
$\hat{\btheta}_k$ to estimate $\btheta_k$.

\section{Bandwidth selection}
\label{BS}

The choice of the bandwidth $h$, used in the estimation of $\bbeta$, is not
crucial.  According to some numerical analysis not presented in 
this paper for brevity, the accuracy of the estimator
$\hat{\bbeta}$ is not very sensitive to $h$, as long as $h$ is within 
in a reasonable range.  In the computational algorithm for 
estimating $\bbeta$, see Section \ref{alg}, we recommend 
choosing a bandwidth $h$ equal to around $20\%$ of the following range
\begin{equation}
\max\{X_{1}^{\T} \tilde{\bbeta}, \cdots, X_{n}^{\T} \tilde{\bbeta}\} -
\min\{X_{1}^{\T} \tilde{\bbeta}, \cdots, X_{n}^{\T} \tilde{\bbeta}\}
\label{range}
\end{equation}
where $\tilde{\bbeta}$ is a randomly chosen initial estimate of $\bbeta$.  We
update $h$ on subsequent iterations by replacing $\tilde{\bbeta}$ in (\ref{range}) with the most recent estimate of ${\bbeta}$. This
approach is employed in the simulation studies and real data analysis of
this paper.

We now focus on the selection of the bandwidth $h_1$, used in the
estimation of $\bg(\cdot)$ and $\bphi(\cdot)$.  The
proposed bandwidth selection is based on a $k$-nearest neighbours bandwidth
with $k$ being selected by cross-validation. We define the cross-validation statistic by
\begin{equation}
\CV(k)
=
\sum\limits_{t=n-M}^n
\left\Vert
Y_{t} - \hat{\bg}^{(t-1)}(X_{t-1}^{\T}\hat{\bbeta})
- \hat{\bphi}^{(t-1)}(X_{t-1}^{\T}\hat{\bbeta}) X_t
\right\Vert
\label{cv}
\end{equation}
where
$\hat{\bg}^{(t-1)}(\cdot)$ and $\hat{\bphi}^{(t-1)}(\cdot)$ are the respective
estimators of $\bg(\cdot)$ and $\bphi(\cdot)$ using a $k$-nearest neighbours
bandwidth based on $(X_l^{\T}, \ Y_l^{\T})$, $l = 1, \ \cdots, \ t-1$, and where $M$ is a look-back integer parameter such that $M<n-1$.

Hence, denoting the $k$ that minimises $\CV(k)$ by $\hat{k}$, we use a $\hat{k}$-nearest neighbours bandwidth in the estimation of $\bg(\cdot)$ and 
$\bphi(\cdot)$.  The bandwidth $h_2$ in the estimation of $\bSigma_x(\cdot)$ or
$E(X_t|X_{t-1} = \bu)$ can also be selected by cross-validation in a similar
way.

\section{Asymptotic properties}
\label{asy}

In this section, we are going to present the asymptotic properties of the
proposed estimators.  We first introduce the following notation which will be used throughout this paper. For any matrix $\bA = (a_{ij})_{m \times N}$, we use
$\lambda_{\min}(\bA)$ and $\lambda_{\max}(\bA)$ to denote respectively the
smallest and largest eigenvalues of $\bA$.  The trace of $\bA$ is denoted by
$\tr(\bA)$, the Frobenius norm of $\bA$ by $\|\bA\|_F$, and the spectral norm (also called operator norm) and element-wise norm by
$$
\|\bA\| = \sqrt{\lambda_{\max}\left(\bA^{\T} \bA\right)},
\quad
\|\bA\|_{\infty}
=
\underset{\underset{{\scriptscriptstyle 1\leq j\leq N}}
{{\scriptscriptstyle 1\leq i\leq m}}}{\mathrm{max}}|a_{ij}|
$$
respectively. We also define
$$
\bU_n
=
\frac{1}{np_n}\sum_{i=2}^{n} \sum_{k =1}^{p_n}
f(X_i^{\T}\bbeta)\{X_{i-1} - \bbE(X_{i}|X_{i-1}^{\T}\bbeta)\}\{\dot{g}_k(X_{i-1}^{\T}\bbeta )
+
\dot{\ba}_{k}(X_{i-1}^{\T}\bbeta) X_{i}\}\epsilon_{k,i}
$$
and
$$
\bV_p
=
{p_n^{-1}}\sum_{k=1}^{p_n}
\bbE \left (f(X_1^{\T}\bbeta)\{\bX_1 - \bbE(\bX_2|\bX_1^{\T}\bbeta)\}^{\otimes 2}
\left\{\dot{g}_k(\bX_1^{\T}\bbeta )
+ \dot{\ba}_{k}(\bX_1^{\T}\bbeta )\bX_{2}\right \}^2 \right ).
$$

\bigskip
\noindent
{\bf Theorem 1.}  {\it Under assumptions (A1 - A5), (B1 - B4), (C1) and
(C3) in Appendix A, there exists $C > 0$ and a small $\varepsilon >0$ such that
\begin{description}
\item {(I)}
$$
\bbP
\left\{
\left\|
\wh{\bbeta} - \bbeta - \bV_p^{-1} \bU_n
\right\|
>
C \left( h^3 + \frac{\log(n ) }{nh} \right)
\right\}
\le
O\left(\frac{1}{n^{1 + \varepsilon}} \right);
$$
\item{(II)}
$$
\bbP
\left\{
\sup_{z \in \mathcal{Z}} \left \|\wh{\bg}(z)  - \bg(z) \right \|_{\infty}
>
C \left( h_1^2 + \sqrt{\frac{\log (n)}{nh_1}}\right)
\right \}
\le O\left({1\over n^{1+ \varepsilon}} \right);
$$
\item {(III)}
$$
\bbP
\left\{
\sup_{z \in \mathcal{Z}} \left \|\wh{\bPhi}(z)  - \bPhi(z)\right \|_{\infty}
>
C \left( h_1^2 + \sqrt{\frac{\log (n)}{nh_1}}\right ) \right\}
\le
O\left({1\over n^{1+ \varepsilon}} \right);
$$
\item {(IV)}
$$
\bbP
\left\{
\sup_{1\le k \le p_n}
\left\|\wh{\btheta}_k - \btheta_k\right\|
>
C \left ( h_1^2 + \sqrt{\frac{\log (n)}{nh_1}} \right)
\right\}
\le
O\left({1\over n^{1 + \varepsilon}}\right),
$$
\end{description}
where $\mathcal{Z}$ is a compact subset of the range of $X_t^{\mathrm{{\scriptstyle T}}} \bbeta$.

}

\bigskip

\noindent
{\bf Remark 1.} Theorem 1 shows that $\|\wh{\bbeta} - \bbeta\| = o_P(n^{-1/2})$
when $p_n$ diverges to $\infty$ as $n \to \infty$, provided that $\|\bU_n\|= o_P(n^{-1/2})$. It indicates that the index $\bbeta$ is estimated with a rate faster than the normal rate $n^{-1/2}$, which is the optimal rate if $p_n$ is fixed. This is known as a `blessing of high dimensionality'.

\bigskip

The main interest of this paper is to estimate $\cov(Y_t | \calF_{t-1})$.  To
measure the accuracy of an estimator $\hat{M}$ of a matrix $M$ of size $p_n$,
we use the entropy loss norm, proposed by James and Stein (1961),
$$
\left\|
\hat{M} - M
\right\|_{\bSigma}
=
p_n^{-1/2}
\left\|
M^{-1/2}
\left\{
\hat{M} - M
\right\}
M^{-1/2}
\right\|_{F}.
$$
To facilitate our presentation, we focus on the convergence of $\wh{\cov}(Y_{n+1}|\mathcal{F}_n) - \cov(Y_{n+1}|\mathcal{F}_n)$, after obtaining the data $\big \{(X_1,Y_1),\cdots,(X_n,Y_n)\big \}$.

\bigskip
\noindent

{\bf Theorem 2.} {\it Under assumptions (A1 - A5), (B1 - B4) and (C1 - C4)
in Appendix A, there exist $C > 0$ and $\varepsilon >0$ such that, 
with probability at least $1 - n^{-(1+\varepsilon)}$, 
\begin{eqnarray*}
\left\|\wh{\mathrm{cov}}(Y_{n+1}| \calF_{n}) - \mathrm{cov}(Y_{n+1} | \calF_{n})
\right\|_{\bSigma}^2
\le
p_n C \left \{ h_1^8 + \left(\frac{\log n}{nh_1} \right )^2\right \}
+
C\left ( h_1^4 + \frac{\log n}{nh_1} \right ) 
+ 
p_n^{-1} C \left ( h_2^4 + \frac{\log n}{nh_2^q} \right ).
\end{eqnarray*} 
}

Fan, Fan and Lv (2008) and Fan, Liao and Mincheva (2011) showed an estimator
of a covariance matrix based on a certain structure would achieve a higher
convergence rate than the sample covariance matrix.  Theorem 2 tells us the
same story.  There are three terms to measure the accuracy of $\wh{\cov}(Y_{n+1}|\mathcal{F}_n) - \cov(Y_{n+1}|\mathcal{F}_n)$. The first two terms 
tell us how the nonparametric smoothing steps in estimating $\bPhi(\cdot)$ affect the performance of $\wh{\cov}(Y_{n+1}|\mathcal{F}_n)$, and the third term evaluates the influence of conditional covariance matrix $\bSigma_{x}(\bX_n)$.  It turns out that even though $q-$dimensional smoothing is required, its effect is small and often negligible if $p_n$ is large.


\section{Computational algorithm}
\label{alg}

To implement the proposed estimation procedure for
$\cov(Y_t|\calF_{t-1})$, the hardest part is to compute an estimate of
$\bbeta$, which is equivalent to finding the minimum of
$$
L(\bg_1, \ \bxi_1, \ A_1, \ B_1, \ \cdots, \ \bg_{n-1}, \ \bxi_{n-1}, \
A_{n-1}, \ B_{n-1}, \ \bbeta)
$$
under the conditions
$$
\|\bbeta \| = 1,
\quad
\beta_1 > 0.
$$
We now introduce the proposed iterative algorithm which can be used to do this minimisation.  Let
\begin{eqnarray*}
&  &
\calQ(\bg_1, \ \bxi_1, \ A_1, \ B_1, \ \cdots, \ \bg_{n-1}, \ \bxi_{n-1}, \
A_{n-1}, \ B_{n-1}, \ \bbeta, \ \bb)
\\
& = &
\sum\limits_{j=1}^{n-1}
\sum\limits_{i=2}^n
\left\|
Y_i - \bg_j - A_j X_i - (\bxi_j + B_j X_i)(X_{i-1} - X_j)^{\T} \bbeta
\right\|^2 K_h((X_{i-1}- X_j)^{\T} \bb),
\end{eqnarray*}
which is
$
L(\bg_1, \ \bxi_1, \ A_1, \ B_1, \ \cdots, \ \bg_{n-1}, \ \bxi_{n-1}, \
A_{n-1}, \ B_{n-1}, \ \bbeta)
$
with the $\bbeta$ in the kernel function being replaced by $\bb$.  First of all, 
randomly choose
an initial estimate for $\bbeta$, denoted by $\tilde{\bbeta}$, such 
that $\|\tilde{\bbeta}\| = 1$ and the first component of $\tilde{\bbeta}$ is 
positive.  Then, iterate between the following two steps until convergence:
\begin{enumerate}
\item[(Step 1)] If this is the first iteration, let $\bbeta_0 = \tilde{\bbeta}$.
Otherwise, set $\bbeta_0$ equal to the $\hat{\bbeta}$ obtained from Step 2 of
the previous iteration.  Minimise
$$
L(\bg_1, \ \bxi_1, \ A_1, \ B_1, \ \cdots, \ \bg_{n-1}, \ \bxi_{n-1}, \
A_{n-1}, \ B_{n-1}, \ \bbeta_0)
$$
with respect to $\bg_1, \ \bxi_1, \ A_1, \ B_1, \ \cdots, \ \bg_{n-1}, \
\bxi_{n-1}, \ A_{n-1}$ and $B_{n-1}$, and denote the minimiser by
$\hat{\bg}_1$, $\hat{\bxi}_1$, $\hat{A}_1$, $\hat{B}_1$, $\cdots$,
$\hat{\bg}_{n-1}$, $\hat{\bxi}_{n-1}$, $\hat{A}_{n-1}$ and $\hat{B}_{n-1}$.

\item[(Step 2)] Minimise
$$
\calQ(\hat{\bg}_1, \ \hat{\bxi}_1, \ \hat{A}_1, \ \hat{B}_1, \ \cdots, \
\hat{\bg}_{n-1}, \ \hat{\bxi}_{n-1}, \ \hat{A}_{n-1}, \ \hat{B}_{n-1}, \
\bbeta, \ \bbeta_0)
$$
with respect to $\bbeta$.  Denote the minimiser by $\check{\bbeta}$, and
define $\hat{\bbeta} = \check{\bbeta}/\|\check{\bbeta}\|$ when the first
component of $\check{\bbeta}$ is positive and $\hat{\bbeta} 
= -\check{\bbeta}/\|\check{\bbeta}\|$ otherwise.
\end{enumerate}
The $\hat{\bbeta}$ resulting from the convergence is the final estimate of $\bbeta$.

The proposed iterative algorithm is easy to implement as both minimisers in Step
1 and Step 2 have a closed form.  Once an estimate of $\bbeta$ is obtained, the remaining
computation of $\cov(Y_t|\calF_{t-1})$ becomes straightforward.

\section{Portfolio allocation}
\label{pa}

In this section, we will briefly describe the construction of an estimated
optimal portfolio allocation based on the proposed dynamic structure and the
associated estimation procedure. Since the formula for optimal portfolio allocation
contains $E(Y_t|\calF_{t-1})$ we shall introduce its estimator
$\wh{E}(Y_t|\calF_{t-1})$ first.
By taking conditional expectation of (\ref{model}), we have
$$
E(Y_t|\calF_{t-1})
=
\bg(X_{t-1}^{\T} \bbeta) + \bphi(X_{t-1}^{\T} \bbeta) E(X_t|X_{t-1}).
$$
Therefore, we use
\begin{equation}
\wh{E}(Y_t|\calF_{t-1})
=
\hat{\bg}(X_{t-1}^{\T} \hat{\bbeta}) + \hat{\bphi}(X_{t-1}^{\T} \hat{\bbeta})
\hat{E}(X_t|X_{t-1})
\label{mean1}
\end{equation}
to estimate $E(Y_t|\calF_{t-1})$ where $\hat{E}(X_t|X_{t-1})$ is
defined in (\ref{mean}).

Our estimated optimal portfolio allocation builds on the mean-variance optimal portfolio 
by Markowitz (1952, 1959). The allocation vector $\bw$ of $p_n$ risky assets, to
be held between times $t-1$ and $t$, is defined as the solution to
\begin{align*}
 & \underset{{\scriptstyle \mathbf{w}}}{\mathrm{min}}\;\mathbf{w}^{\mathrm{
 {\scriptstyle T}}}\mathrm{cov}(Y_{t}|\mathcal{F}_{t-1})\mathbf{w}\\
 & \mbox{subject to }\mathbf{w}^{\mathrm{{\scriptstyle T}}}
 \mathbf{1}_{p_n}=1\quad\mbox{and}\quad\mathbf{w}^{\mathrm{
 {\scriptstyle T}}}E(Y_{t}|\mathcal{F}_{t-1})=\delta
\end{align*}
where $\delta$ is the target return imposed on the portfolio. 
The solution $\hat{\bw}$ is given by
\begin{equation}
\hat{\bw}
=
\frac{c_3 - c_2 \delta}{c_1 c_3 - c_2^2}
\wh{\cov}(Y_t|\calF_{t-1})^{-1} \1_{p_n}
+
\frac{c_1 \delta - c_2}{c_1 c_3 - c_2^2}
\wh{\cov}(Y_t|\calF_{t-1})^{-1} \wh{E}(Y_t|\calF_{t-1}),
\label{por}
\end{equation}
where
$$
c_1
=
\1_{p_n}^{\T} \wh{\cov}(Y_t|\calF_{t-1})^{-1} \1_{p_n},
\quad
c_2
=
\1_{p_n}^{\T} \wh{\cov}(Y_t|\calF_{t-1})^{-1} \wh{E}(Y_t|\calF_{t-1}),
$$
$$
c_3
=
\wh{E}(Y_t|\calF_{t-1})^{\T}
\wh{\cov}(Y_t|\calF_{t-1})^{-1}
\wh{E}(Y_t|\calF_{t-1}).
$$

\section{Simulation studies}
\label{sim}

In this section, we are going to use a simulated example to show how well
the proposed estimation procedure and portfolio allocation works.  We 
shall use $a_{i,j}(\cdot)$ to denote the entry corresponding to the $i$th row and $j$th column of $\bphi(\cdot)$.

We generate 1000 data sets from
model (\ref{model}) together with (\ref{gar}).
We repeat this using the following combinations 
of $n$ and $p_n$: 
$\{n=1000, p_n=50\}$, 
$\{n=1000, p_n=100\}$, 
$\{n=2000, p_n=50\}$ and 
$\{n=2000, p_n=100\}$.
We set
$$
q = 4,
\quad
m=1,
\quad
s=1,
\quad
\bbeta
=
\frac{1}{3}(1, \ 2, \ 0, \ 2)^{\T}.
$$
For $k=1, \ \cdots, \ p_{n}$, we set
$$
\alpha_{0,k}=0.5,
\quad
\alpha_{1,k}=0.1,
\quad
\beta_{1,k}=0.1,
\quad
g_{k}(z)
=
\Xi_{0,k}+3\mbox{exp}(-z^{2}),
\quad
a_{k,1}(z)
=
\Xi_{1,k}+0.8z,
$$
$$
a_{k,2}(z)
=
\Xi_{2,k},
\quad
a_{k,3}(z)
=
\Xi_{3,k}+1.5\mbox{sin}(\pi z),
\quad
a_{k,4}(z)
=
\Xi_{4,k},
$$
where $\Xi_{j,k}$ are some fixed parameters for
$j=0, \ \cdots, \ d$ and $k=1, \ \cdots, \ p_{n}$.  In order to define
$\Xi_{j,k}$, we simulate them independently from a uniform distribution on
$[-1, \ 1]$, and use these same values throughout all simulations.
For $t = 1, \ \cdots, \ n+1$, we generate $X_t$ independently from a uniform
distribution on $[-1,1]^q$, $Z_t$ from
$p_n$-variate standard normal distribution, and $\ep_t$
through $\ep_t = \bSigma_{0,t}^{1/2} Z_t$.  Once both $X_t$ and $\ep_t$
have been generated, $Y_t$ can be generated through (\ref{model}) for 
$t = 1, \ \cdots, \ n+1$.

We will initially pretend that $(X_{n+1}^{\T}, \ Y_{n+1}^{\T})$ is unknown to us, and
this will not be used in the estimation of $\cov(Y_{n+1}|\calF_n)$.  The purpose of generating an
additional data point $(X_{n+1}^{\T}, \ Y_{n+1}^{\T})$ is to enable us to
calculate the 1-period simple return 
\begin{equation}
R(\hat{\bw}) = \hat{\bw}^{\T} Y_{n+1}
\label{re}
\end{equation}
of a portfolio allocation $\hat{\bw}$ formed at time
$n$ based on data $(X_t^{\T}, \ Y_t^{\T})$, $t= 1, \ \cdots, \ n$.
In order to evaluate the performance of an estimator $\hat{M}$ of matrix $M$
we use the following metric
\[
\Delta(\hat{M}, \ M)
=
\dfrac{
\bigl\Vert\hat{M}-M\bigr\Vert_F
}
{\bigl\Vert M\bigr\Vert_F}.
\]
We also use the Sharpe ratio
$$
\SR(\hat{\bw})
=
\frac{E\left\{R(\hat{\bw})\right\}}
{\SD\left\{R(\hat{\bw})\right\}}
$$
to evaluate the performance of $\hat{\bw}$, where
$\SD\left\{R(\hat{\bw})\right\}$ is the standard deviation of
$R(\hat{\bw}).$ We assume a zero risk-free rate for simplicity.

We first examine how well the estimation procedure works. We estimate
$\cov(Y_{n+1}|\calF_n)$, and use $\wh{\cov}(Y_{n+1}|\calF_n)^{-1}$ to estimate
$\cov(Y_{n+1}|\calF_n)^{-1}$.  
The kernel function in the estimation procedure
is taken to be the Epanechnikov kernel $K(u) = 0.75(1 - u^2)_+$, and the
bandwidths are selected by the methodology described in Section \ref{BS}.
The results, presented in Tables \ref{tab1} and \ref{tab2}, show both $\wh{\cov}(Y_{n+1}|\calF_n)$  and $\wh{\cov}(Y_{n+1}|\calF_n)^{-1}$ work very well.
\begin{table}[htbp]
\begin{center}
\caption{{\bf Mean and Standard Deviation of
$\Delta\left(
\wh{\mathrm{cov}}(Y_{n+1}|\calF_n), \ \mathrm{cov}(Y_{n+1}|\calF_n)
\right)
$}}
\label{tab1}
\vspace{0.3cm}
\begin{tabular}{c|cccc}
\hline\hline
 &
$
\begin{array}{c}
n = 1000
\\
p_n = 50
\end{array}
$
&
$
\begin{array}{c}
n = 1000
\\
p_n = 100
\end{array}
$
&
$
\begin{array}{c}
n = 2000
\\
p_n = 50
\end{array}
$
&
$
\begin{array}{c}
n = 2000
\\
p_n = 100
\end{array}
$
\\\hline
$E(D)$ & 0.183 & 0.189 & 0.136 & 0.141
\\\hline
$\SD(D)$ & 0.046 & 0.049 & 0.034 & 0.035
\\\hline\hline
\end{tabular}
\end{center}
\begin{singlespace}
\begin{center}
\begin{minipage}{30pc}
{\it In this table,
$D
=
\Delta
\left(
\wh{\mathrm{cov}}(Y_{n+1}|\calF_n), \ \mathrm{cov}(Y_{n+1}|\calF_n)
\right)$, and
$\mathrm{SD}(D)$ is the standard deviation of $D$.}
\end{minipage}
\end{center}
\end{singlespace}
\end{table}
\begin{table}[htbp]
\begin{center}
\caption{{\bf
Mean and Standard Deviation of
$\Delta\left(
\wh{\mathrm{cov}}(Y_{n+1}|\calF_n)^{-1}, \ \mathrm{cov}(Y_{n+1}|\calF_n)^{-1}
\right)
$
}}
\label{tab2}
\vspace{0.3cm}
\begin{tabular}{c|cccc}
\hline\hline
 &
$
\begin{array}{c}
n = 1000
\\
p_n = 50
\end{array}
$
&
$
\begin{array}{c}
n = 1000
\\
p_n = 100
\end{array}
$
&
$
\begin{array}{c}
n = 2000
\\
p_n = 50
\end{array}
$
&
$
\begin{array}{c}
n = 2000
\\
p_n = 100
\end{array}
$
\\\hline
$E(D_1)$ & 0.114 & 0.105 & 0.078 & 0.070
\\\hline
$\SD(D_1)$ & 0.017 & 0.013 & 0.012 & 0.009
\\\hline\hline
\end{tabular}
\end{center}
\begin{singlespace}
\begin{center}
\begin{minipage}{30pc}
{\it In this table,
$D_1
=
\Delta
\left(
\wh{\mathrm{cov}}(Y_{n+1}|\calF_n)^{-1}, \ \mathrm{cov}(Y_{n+1}|\calF_n)^{-1}
\right)$, and
$\mathrm{SD}(D_1)$ is the standard deviation of $D_1$.}
\end{minipage}
\end{center}
\end{singlespace}
\end{table}

We now examine the performance of the proposed portfolio allocation, using a target return $\delta=1\%$, by computing the return as described in (\ref{re}). In order to see 
how much gain can be made by making use of the dynamic structure, we make a comparison
with portfolio allocations based on Markowitz's formula but where the covariance matrix is 
estimated using the sample covariance matrix and also the estimator proposed 
by Fan, Fan and Lv (2008). The mean, standard deviation and Sharpe ratio of
the returns are presented in Table \ref{tab3}. For each situation discussed, 
we see the Sharpe ratio of the
proposed portfolio allocation is much bigger than the other
two portfolio allocations.  This suggests there is significant gain from making use of the
dynamic structure of the covariance matrix.

\begingroup
\begin{table}[htbp]
\renewcommand*{\arraystretch}{0.8}

\begin{center}
\caption{{\bf Means, Standard Deviations and Sharpe Ratios }}
\label{tab3}
\vspace{0.3cm}
\begin{tabular}{c|cccc}
\hline\hline
 &
$
\begin{array}{c}
n = 1000
\\
p_n = 50
\end{array}
$
&
$
\begin{array}{c}
n = 1000
\\
p_n = 100
\end{array}
$
&
$
\begin{array}{c}
n = 2000
\\
p_n = 50
\end{array}
$
&
$
\begin{array}{c}
n = 2000
\\
p_n = 100
\end{array}
$
\\\hline
$E\left\{R(\hat{\bw})\right\}$ & 0.99\% & 1.01\% & 1.03\% & 1.03\%
\\
$E\left\{R(\hat{\bw}_1)\right\}$ & 0.96\% & 0.96\% & 1.02\% & 1.02\%
\\
$E\left\{R(\hat{\bw}_2)\right\}$ & 0.96\% & 0.96\% & 1.02\% & 1.02\%
\\\hline
$\SD\left\{R(\hat{\bw})\right\}$ & 0.40\% & 0.28\% & 0.39\% & 0.27\%
\\
$\SD\left\{R(\hat{\bw}_1)\right\}$ & 1.02\% & 1.03\% & 1.03\% & 1.02\%
\\
$\SD\left\{R(\hat{\bw}_2)\right\}$ & 0.99\% & 0.97\% & 1.02\% & 1.00\%
\\\hline
$\SR(\hat{\bw})$ & 2.49 & 3.57 & 2.63 & 3.83
\\
$\SR(\hat{\bw}_1)$ & 0.94 & 0.93 & 0.99 & 1.00
\\
$\SR(\hat{\bw}_2)$ & 0.97 & 0.99 & 1.00 & 1.02
\\\hline\hline
\end{tabular}
\end{center}
\begin{singlespace}
\begin{center}
\begin{minipage}{30pc}
{\it In this table we denote the proposed portfolio allocation by $\hat{\bw}$,
the portfolio allocation formed by Markowitz's formula using the 
sample covariance matrix by $\hat{\bw}_1$, and the portfolio allocation formed by Markowitz's formula using the estimated covariance matrix from Fan, Fan and Lv
(2008) by $\hat{\bw}_2$. }
\end{minipage}
\end{center}
\end{singlespace}
\end{table}

\endgroup

\section{Real data analysis}
\label{real}

In this section, we are going to apply the dynamic structure for
covariance matrices to a real data set.  We use the term {\it Face} (Factor model with an
Adaptive-varying-coefficient-model structure
Covariance matrix Estimator) to denote the proposed
portfolio allocation.  This name was chosen because the estimator 
will `face' the markets today based on what happened yesterday and adapt according
to the dynamic structure.
We compare Face with the allocation based on the sample covariance matrix 
(denoted by {\it Sam}), and the allocation proposed by Fan, Fan and 
Lv (2008) (denoted by {\it Fan}).  In all three cases, we use the same target return
$\delta = 1\%$.  We also make a comparison with the market portfolio 
(denoted by {\it Market}) since this aids as an important benchmark  
indicating whether we are in a bull or bear market. In this section, 
the kernel function used in the construction of Face is still taken
to be the Epanechnikov kernel, and the bandwidths are selected by the method
described in Section \ref{BS}.

All data used can be freely downloaded from Kenneth French's website
\url{http://mba.tuck.dartmouth.edu/pages/faculty/ken.french/data_library.html}
and was accessed on 2nd April 2015. The response variable $Y_t$ is chosen to be the
vector of the daily returns of $p_n = 49$ industry portfolios (value weighted) minus the
risk-free rate. The
observable factors $x_{1,t}$, $x_{2,t}$ and $x_{3,t}$ are taken to be the
market, size and value factors respectively from the Fama-French three-factor model.
The labelling along with a brief description of
$Y_t = (y_{1,t}, \ \cdots, y_{49,t})^{\T}$ and
$X_t = (x_{1,t}, \ x_{2,t}, \ x_{3,t})^{\T}$ can be found in
Table \ref{tab4} and Table \ref{tab5} respectively.

There are various advantages of using the portfolio returns for $y_{k,t}$ as
opposed to using individual stocks: we avoid having to merge different sources
of data; we avoid survivorship bias (where we only picked companies that did
not go bankrupt); and we attempt to avoid company specific risk.  A further
benefit is that the results we give are entirely reproducible
since the data is free and presented in a spreadsheet format.

\begin{figure}[H]
\caption{
\label{fig1}
{\bf  Returns plots of factors and the risk-free rate $R_{f}$}}

\begin{centering}
\includegraphics[width=5.5in]{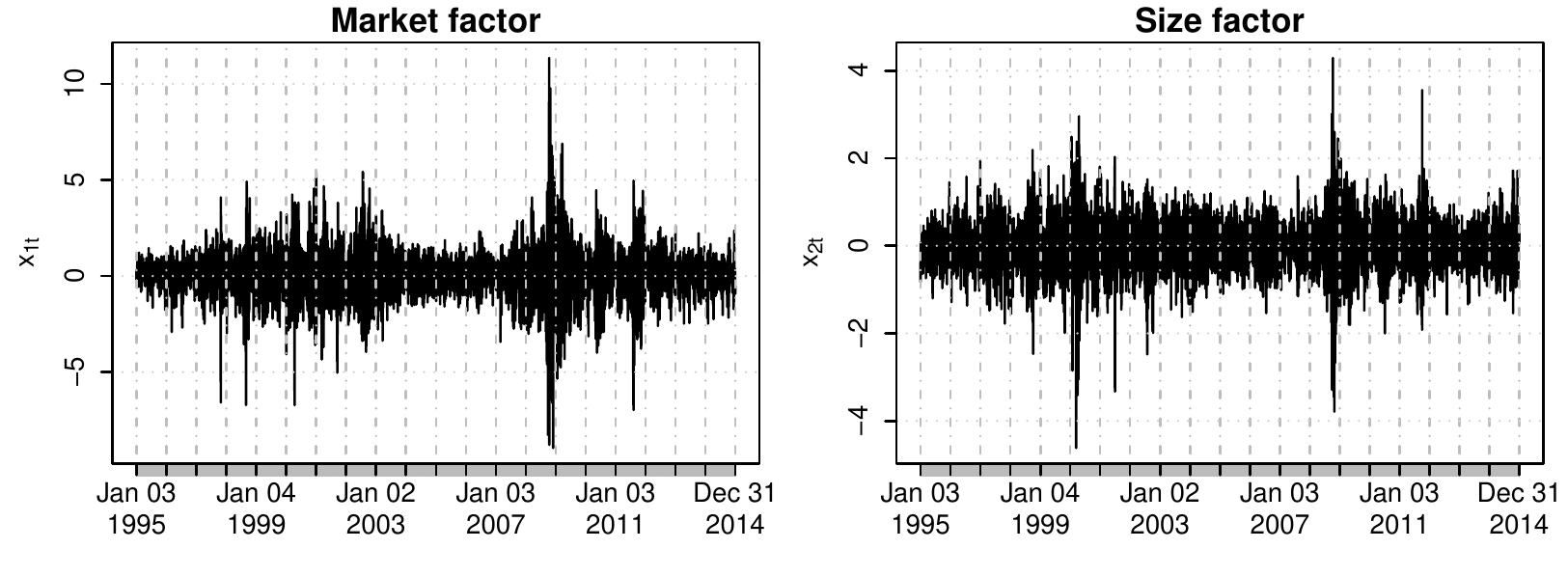}
\par\end{centering}

\begin{centering}
\includegraphics[width=5.5in]{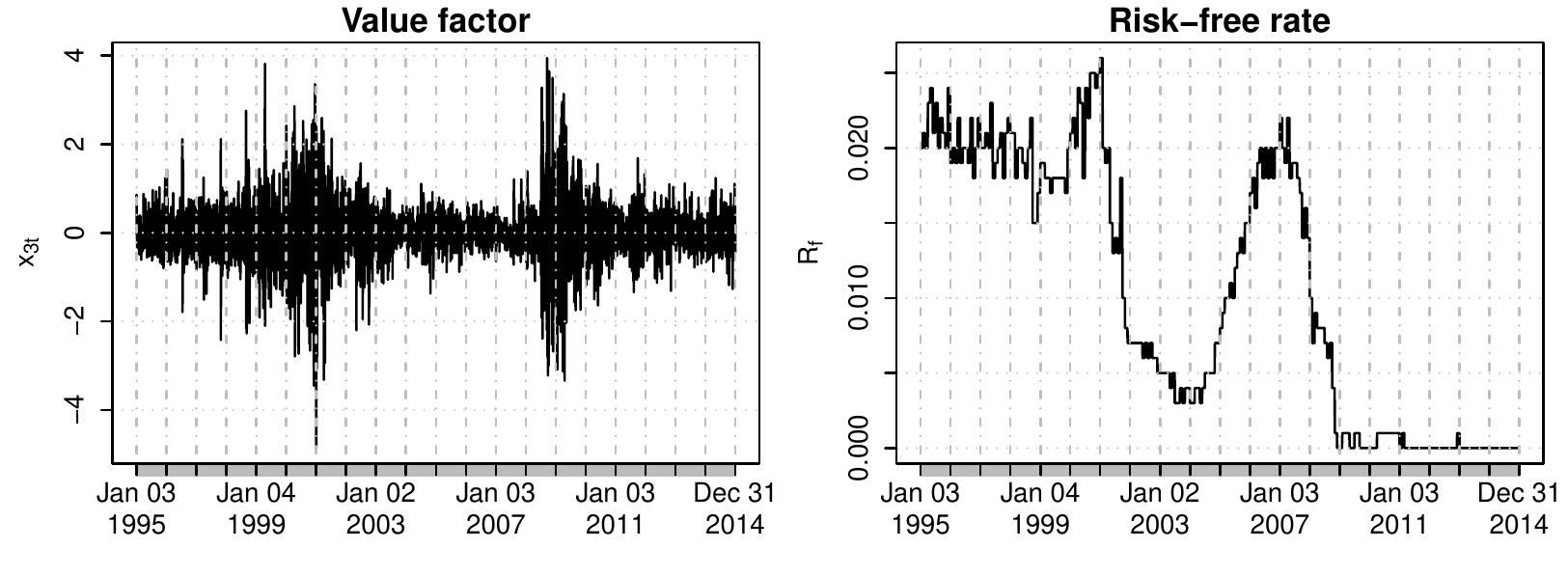}
\par\end{centering}
\end{figure}

\begin{figure}[H]
\caption{
\label{fig2}
{\bf Returns plots of $y_{1,t}$, $y_{2,t}$, $y_{3,t}$, and $y_{4,t}$. }}

\begin{centering}
\includegraphics[width=5.5in]{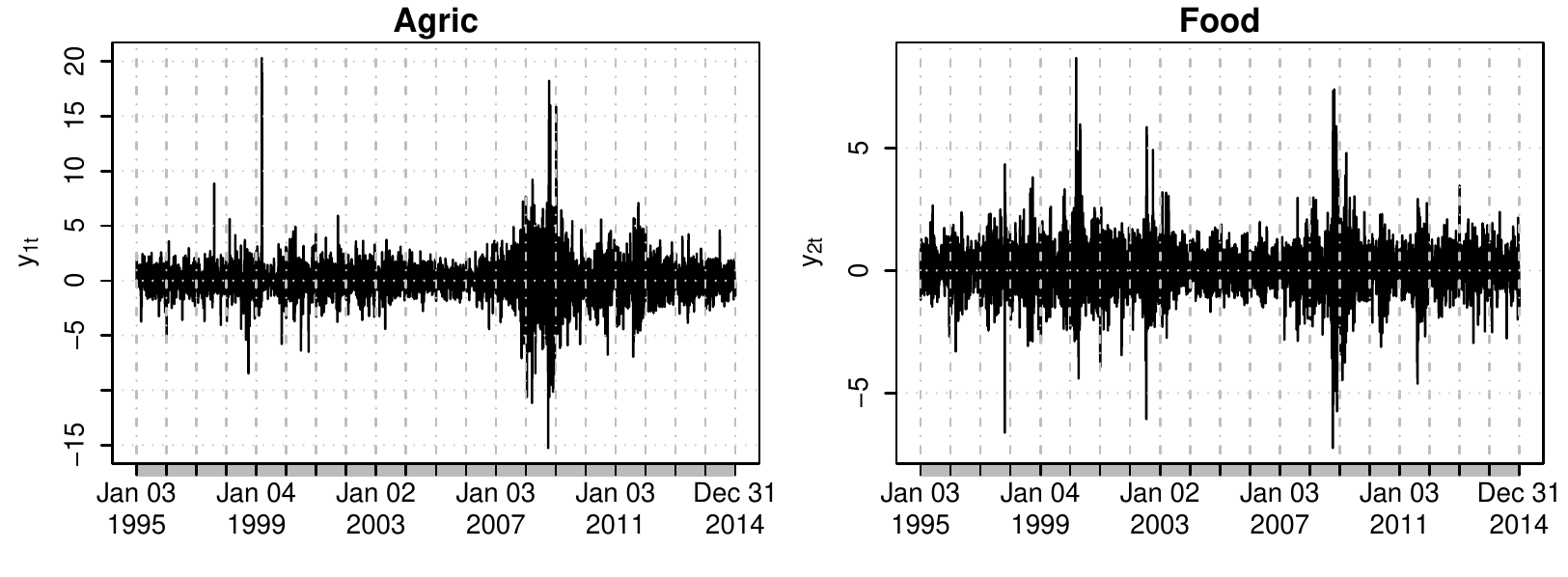}
\par\end{centering}

\begin{centering}
\includegraphics[width=5.5in]{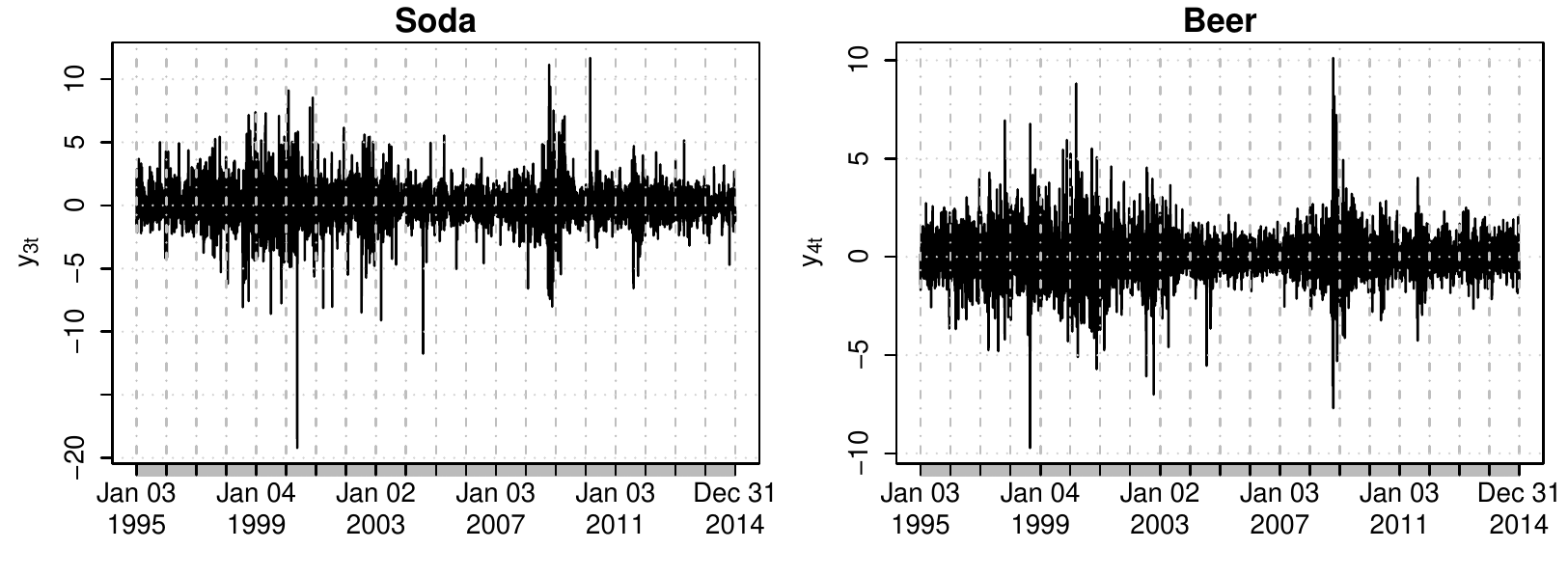}
\par\end{centering}

\end{figure}

To have a better idea about what the data is like, we plot the observations
from 3rd January 1995 to 31st December 2014 of
the three factors and the risk-free rate in Figure \ref{fig1}, and the first four components of $Y_t$ in Figure \ref{fig2}
corresponding to the industrial sectors Agriculture, Food Products, Candy \& Soda, and
Beer \& Liquor.  The plots show
clearly that there are periods of large volatility around the 2008-2009
financial crisis.  We will see Face performs reasonably well even during that
period, whilst the others do not.

We compare the three portfolio allocations, (Face, Sam and Fan), along
with the market portfolio, year by year from 1995 to 2014 using a simple
trading strategy.  For each year we trade
on each trading day, which is approximately $T = 252$ trading days per year.  At the
beginning of each year we assume we have an initial balance of $100$ pounds.
Although this initial choice is arbitrary, it is a useful way of comparing
the performance during the course of a year.  We assume no transaction costs,
allow for short selling, and assume that all possible portfolio allocations 
are attainable.  Our trading strategy consists of forming a portfolio allocation
$\hat{\bw}$ the end of each trading day and holding it until the end of the
next trading day.  Between 
day $t - 1$ and day $t$, we obtain the portfolio return
$$
R_t(\hat{\bw}) = \hat{\bw}^{\T} Y_t
$$
where $\hat{\bw}$ is formed based on $(X_{t-j}^{\T}, \ Y_{t-j}^{\T})$, $j=1, \
\cdots, \ n$, for some look-back integer $n$.  With the realised returns
$R_t(\hat{\bw})$, $t = 1, \ \cdots, \ T$, we can calculate the annualized
Sharpe ratio
$$
\SR(\hat{\bw})
=
\frac{\bar{R}(\hat{\bw})}{SD(R)} \sqrt{T},
$$
where
$$
\bar{R}(\hat{\bw})
=
\frac{1}{T} \sum\limits_{t=1}^T
\left\{
R_t(\hat{\bw}) - R_{f,t}
\right\},
\quad
SD(R)
=
\left[
\frac{1}{T} \sum\limits_{t=1}^T
\left\{
R_t(\hat{\bw}) - R_{f,t} - \bar{R}(\hat{\bw})
\right\}^2
\right]^{1/2}
$$
and $R_{f,t}$ is the risk-free rate on day $t$.
Hence, for each year, and for each of the four trading strategies, we
compute an annualized Sharpe ratio and the balance at the end of the 
final trading day of the year. We repeat this using $n=100$, $300$, and $500$.
From the the annualized Sharpe ratios presented 
in Figure \ref{fig3} and the balances in 
Table \ref{tab6}, it is clear that Face performs significantly better than 
the other three.

We remark that although Face, Sam and Fan are all constructed based on Markowitz's 
formula, the difference between them lies in 
the way to estimate the covariance matrix of
returns, which appears in Markowitz's formula.  
Both Sam and Fan do not take
into account the dynamic feature of the covariance matrix in their estimation,
but Face does. This is the fundamental reason why Face performs significantly better than Sam and Fan.  One may argue that if Sam and Fan used fewer observations in their moving window 
to estimate the covariance matrix they would start to take the dynamic feature
into account, potentially improving their performance.  However when constructing Face, Sam and Fan,
we tried a variety of $n$, ranging from $100$ to $500$,  and found Face
always performs better.  This suggests that even if Sam and Fan only use the
observations in a carefully chosen moving window, Face still outperforms them.

\begin{figure}[H]
\begin{centering}
\caption{
\label{fig5}
{\bf Estimated coefficient functions for industry portfolios 1-4}
}
\par\end{centering}
\vspace{0.1cm}

\begin{centering}
\includegraphics[width=5in]{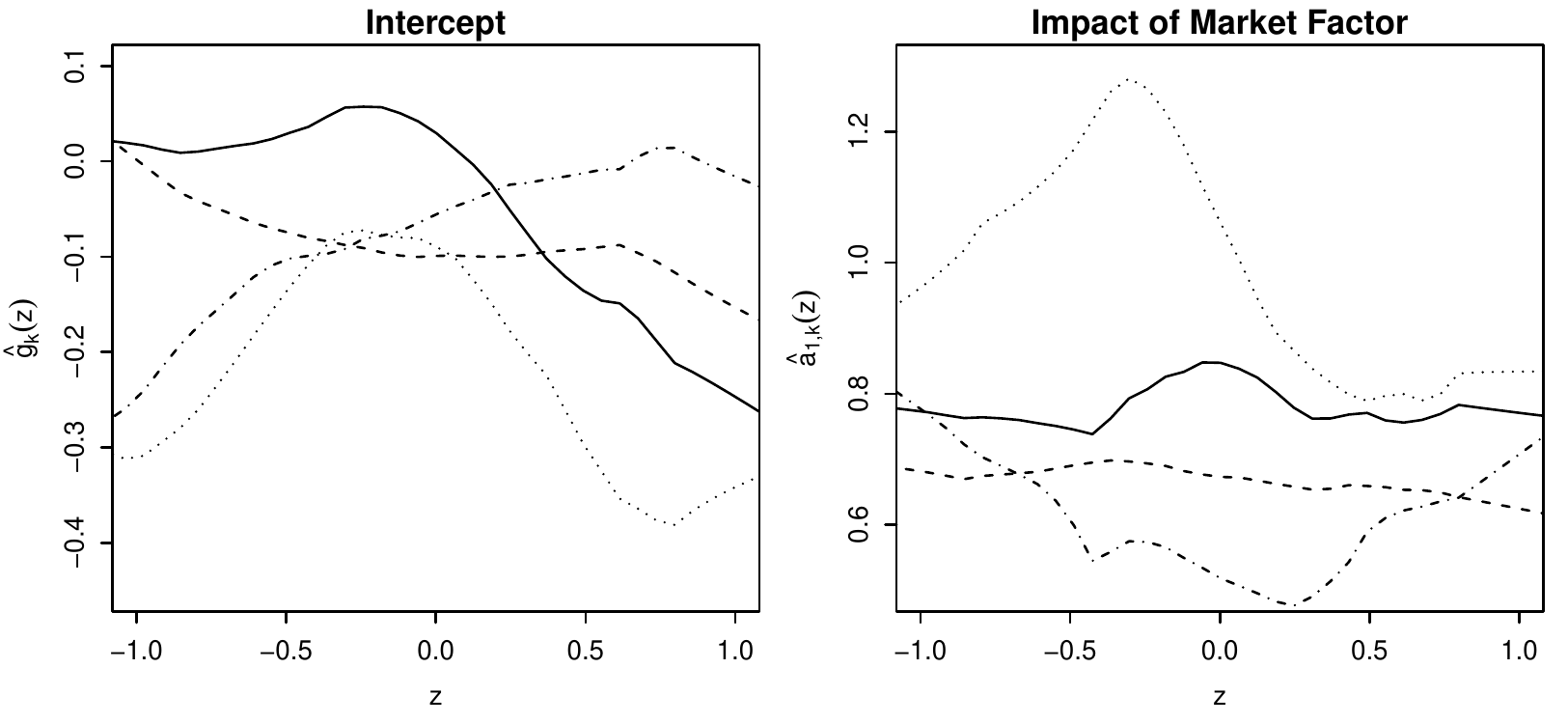}
\par\end{centering}

\begin{centering}
\includegraphics[width=5in]{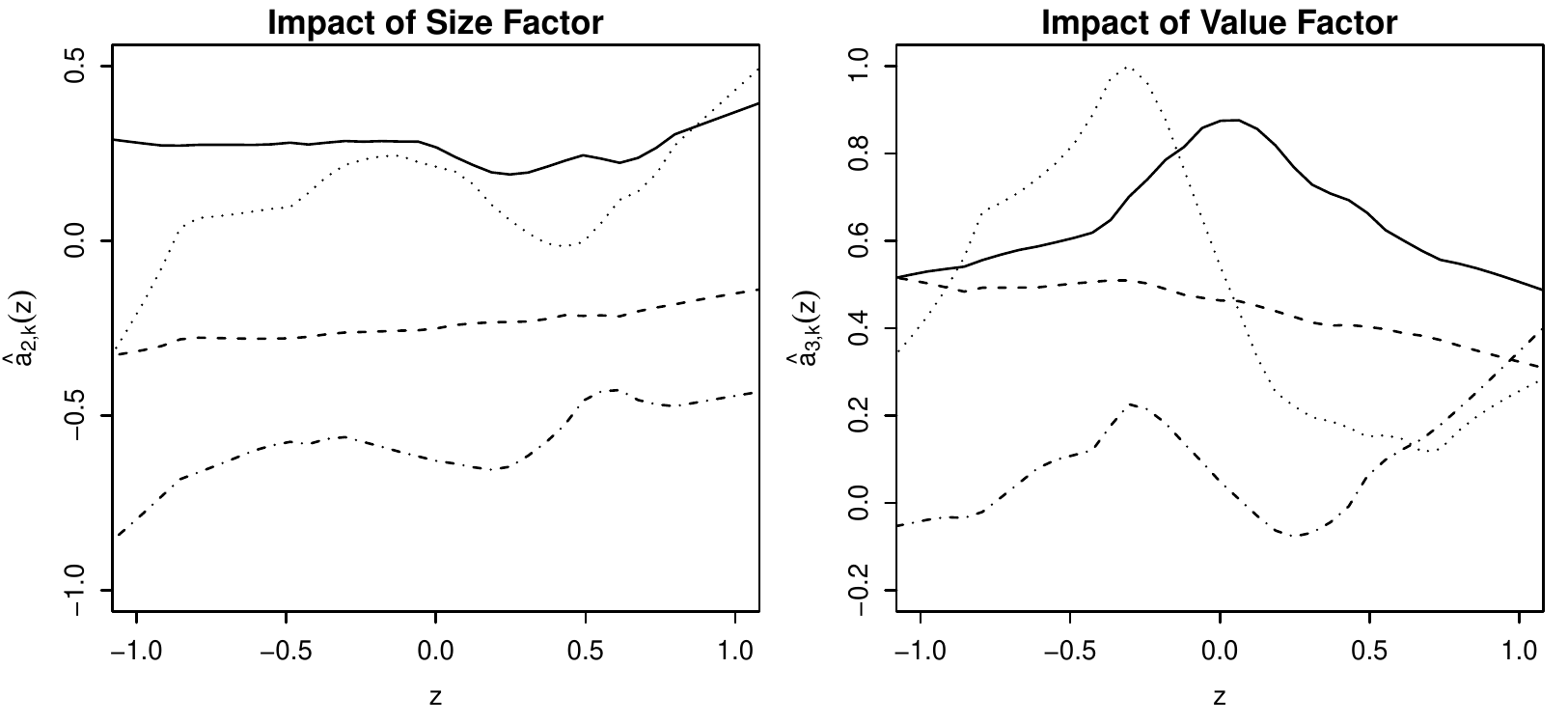}
\par\end{centering}

\vspace{0.1cm}

\centering{}\includegraphics[width=5in]{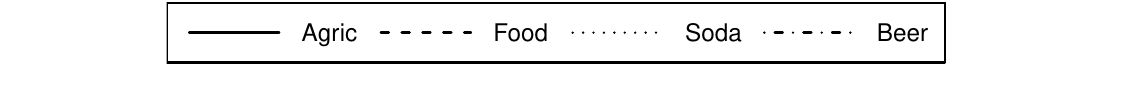}
\begin{singlespace}
\begin{center}
\begin{minipage}{30pc}
{\it This figure shows the estimated intercept and coefficient functions 
for the market, size and value factors, for the first four industry portfolios
(Agriculture, Food Products, Candy \& Soda, and Beer \& Liquor) on the first day of trading.}
\end{minipage}
\end{center}
\end{singlespace}

\end{figure}

To have a tangible idea about whether the covariance matrix is dynamic or not,
we plot the estimated intercept and coefficients of $x_{1,t}$, $x_{2,t}$
and $x_{3,t}$, interpreted as the impact of the factors, for each of the first four components of $Y_t$ in Figure \ref{fig5}.
One can see that these coefficients are dynamic rather than constant, 
which implies the covariance matrix is also dynamic.

It is interesting to have a closer look at the performances of the four strategies
in the volatile time period 2007-2009 during which the financial crisis
took place. Still assuming an initial balance of 100 pounds at the start of each year,
and using $n=500$, we plot the balances at the end of each trading day in Figure \ref{fig4}. 
During 2007, Face, Sam and Fan all perform reasonably well, with Face slightly better. 
The market does not make much profit, and is beaten by the other 
three.  In 2008, Face continuously does
well whilst the other three do not make profit at all.  In 2009, although Face does not do very well
during some time periods, it adapts to the market change quickly and almost
breaks even.  The reason that Face can adapt to market change quickly is because it
takes into account the dynamic feature of the covariance matrix of 
returns.  On the other hand, both Sam and Fan do very poorly, and in fact they
almost lose all their money at the end of the year.  In 2009, the market performs best,
but still with very little profit.

\begingroup
\begin{table}[H]
\renewcommand*{\arraystretch}{0.65}
\begin{centering}
\caption{\label{tab4} {\bf Description of the 49 industry
portfolios}}
\par\end{centering}
\centering{}{\footnotesize{}}%
\begin{tabular}{cc>{\centering}p{5cm}ccc>{\centering}p{5cm}}
\hline
\hline
$k$ & $y_{k,t}$ & Industry name &  & $k$ & $y_{k,t}$ &
Industry name\tabularnewline[1mm]
{\footnotesize{1}} & \texttt{\footnotesize{Agric}} &
{\footnotesize{Agriculture }} &  & {\footnotesize{26}} &
\texttt{\footnotesize{Guns}} & {\footnotesize{Defense }}
\tabularnewline
{\footnotesize{2}} & \texttt{\footnotesize{Food}} &
{\footnotesize{Food Products }} &  & {\footnotesize{27}} &
\texttt{\footnotesize{Gold}} & {\footnotesize{Precious Metals }}
\tabularnewline
{\footnotesize{3}} & \texttt{\footnotesize{Soda}} &
{\footnotesize{Candy \& Soda }} &  & {\footnotesize{28}} &
\texttt{\footnotesize{Mines}} &
{\footnotesize{Industrial Metal Mining }}
\tabularnewline
{\footnotesize{4}} & \texttt{\footnotesize{Beer}} &
{\footnotesize{Beer \& Liquor }} &  & {\footnotesize{29}} &
\texttt{\footnotesize{Coal}} & {\footnotesize{Coal }}
\tabularnewline
{\footnotesize{5}} & \texttt{\footnotesize{Smoke}} &
{\footnotesize{Tobacco Products }} &  & {\footnotesize{30}} &
\texttt{\footnotesize{Oil}} & {\footnotesize{Petroleum and Natural Gas }}
\tabularnewline
{\footnotesize{6}} & \texttt{\footnotesize{Toys}} &
{\footnotesize{Recreation }} &  & {\footnotesize{31}} &
\texttt{\footnotesize{Util}} & {\footnotesize{Utilities }}
\tabularnewline
{\footnotesize{7}} & \texttt{\footnotesize{Fun}} &
{\footnotesize{Entertainment }} &  & {\footnotesize{32}} &
\texttt{\footnotesize{Telcm}} & {\footnotesize{Communication }}
\tabularnewline
{\footnotesize{8}} & \texttt{\footnotesize{Books}} &
{\footnotesize{Printing and Publishing }} &  & {\footnotesize{33}} &
\texttt{\footnotesize{PerSv}} & {\footnotesize{Personal Services }}
\tabularnewline
{\footnotesize{9}} & \texttt{\footnotesize{Hshld}} &
{\footnotesize{Consumer Goods }} &  & {\footnotesize{34}} &
\texttt{\footnotesize{BusSv}} & {\footnotesize{Business Services }}
\tabularnewline
{\footnotesize{10}} & \texttt{\footnotesize{Clths}} &
{\footnotesize{Apparel }} &  & {\footnotesize{35}} &
\texttt{\footnotesize{Hardw}} & {\footnotesize{Computers }}
\tabularnewline
{\footnotesize{11}} & \texttt{\footnotesize{Hlth}} &
{\footnotesize{Healthcare }} &  & {\footnotesize{36}} &
\texttt{\footnotesize{Softw}} & {\footnotesize{Computer Software }}
\tabularnewline
{\footnotesize{12}} & \texttt{\footnotesize{MedEq}} &
{\footnotesize{Medical Equipment }} &  & {\footnotesize{37}} &
\texttt{\footnotesize{Chips}} & {\footnotesize{Electronic Equipment }}
\tabularnewline
{\footnotesize{13}} & \texttt{\footnotesize{Drugs}} &
{\footnotesize{Pharmaceutical Products }} &  & {\footnotesize{38}} &
\texttt{\footnotesize{LabEq}} &
{\footnotesize{Measuring and Control Equipment }}
\tabularnewline
{\footnotesize{14}} & \texttt{\footnotesize{Chems}} &
{\footnotesize{Chemicals }} &  & {\footnotesize{39}} &
\texttt{\footnotesize{Paper}} & {\footnotesize{Business Supplies }}
\tabularnewline
{\footnotesize{15}} & \texttt{\footnotesize{Rubbr}} &
{\footnotesize{Rubber and Plastic Products }} &  & {\footnotesize{40}} &
\texttt{\footnotesize{Boxes}} & {\footnotesize{Shipping Containers }}
\tabularnewline
{\footnotesize{16}} & \texttt{\footnotesize{Txtls}} &
{\footnotesize{Textiles }} &  & {\footnotesize{41}} &
\texttt{\footnotesize{Trans}} & {\footnotesize{Transportation }}
\tabularnewline
{\footnotesize{17}} & \texttt{\footnotesize{BldMt}} &
{\footnotesize{Construction Materials }} &  & {\footnotesize{42}} &
\texttt{\footnotesize{Whlsl}} & {\footnotesize{Wholesale }}
\tabularnewline
{\footnotesize{18}} & \texttt{\footnotesize{Cnstr}} &
{\footnotesize{Construction }} &  & {\footnotesize{43}} &
\texttt{\footnotesize{Rtail}} & {\footnotesize{Retail }}
\tabularnewline
{\footnotesize{19}} & \texttt{\footnotesize{Steel}} &
{\footnotesize{Steel Works Etc }} &  & {\footnotesize{44}} &
\texttt{\footnotesize{Meals}} & {\footnotesize{Restaurants, Hotels, Motels }}
\tabularnewline
{\footnotesize{20}} & \texttt{\footnotesize{FabPr}} &
{\footnotesize{Fabricated Products }} &  & {\footnotesize{45}} &
\texttt{\footnotesize{Banks}} & {\footnotesize{Banking }}
\tabularnewline
{\footnotesize{21}} & \texttt{\footnotesize{Mach}} &
{\footnotesize{Machinery }} &  & {\footnotesize{46}} &
\texttt{\footnotesize{Insur}} & {\footnotesize{Insurance }}
\tabularnewline
{\footnotesize{22}} & \texttt{\footnotesize{ElcEq}} &
{\footnotesize{Electrical Equipment }} &  & {\footnotesize{47}} &
\texttt{\footnotesize{RlEst}} & {\footnotesize{Real Estate }}
\tabularnewline
{\footnotesize{23}} & \texttt{\footnotesize{Autos}} &
{\footnotesize{Automobiles and Trucks }} &  & {\footnotesize{48}} &
\texttt{\footnotesize{Fin}} & {\footnotesize{Trading }}
\tabularnewline
{\footnotesize{24}} & \texttt{\footnotesize{Aero}} &
{\footnotesize{Aircraft }} &  & {\footnotesize{49}} &
\texttt{\footnotesize{Other}} & {\footnotesize{Almost Nothing }}
\tabularnewline
{\footnotesize{25}} & \texttt{\footnotesize{Ships}} &
{\footnotesize{Shipbuilding, Railroad Equipment }} &  &  &  &
\tabularnewline
\hline
\hline
\end{tabular}
\end{table}
\endgroup
\begin{singlespace}
\begin{center}
\begin{minipage}{30pc}
{\it This table gives the labelling and a brief description of industrial sectors which form the 49 Industry Portfolios data set. Precise details of their construction are given on Kenneth French's website.}
\end{minipage}
\end{center}
\end{singlespace}
\emph{}

\begingroup
\renewcommand*{\arraystretch}{0.65}
\begin{table}[H]
\caption{\label{tab5}
{\bf Description of the Fama and French
factors}}
\vspace{0.1cm}
\centering{}{\footnotesize{}}%

\begin{tabular}{c>{\centering}p{1in}>{\centering}p{3in}}
\hline
\hline
\noalign{\vskip1mm}
$j$ & Name of $x_{j,t}$ & Description\tabularnewline[1mm]
{\footnotesize{1}} & {\footnotesize{Market factor}}  &
{\footnotesize{Return on the market minus the risk-free rate}}
\tabularnewline[1mm]

{\footnotesize{2}}  & {\footnotesize{Size factor}} &
{\footnotesize{Excess returns of small caps over big caps}}
\tabularnewline[1mm]

{\footnotesize{3}} & {\footnotesize{Value factor}} &
{\footnotesize{Excess returns of value stocks over growth stocks}}
\tabularnewline[1mm]
\hline
\hline
\end{tabular}
\end{table}

\begin{singlespace}
\begin{center}
\begin{minipage}{30pc}
{\it This table gives the labelling and a brief description of market, size and value factors from the Fama-French factors data set. Precise details of their construction are given on Kenneth French's website.}
\end{minipage}
\end{center}
\end{singlespace}
\endgroup

\vspace{-0.75cm}
\emph{}

\begin{center}
\begin{figure}[H]
\caption{
\label{fig3}
{\bf Annualized Sharpe Ratios}}

\begin{centering}
\includegraphics[width=5in]{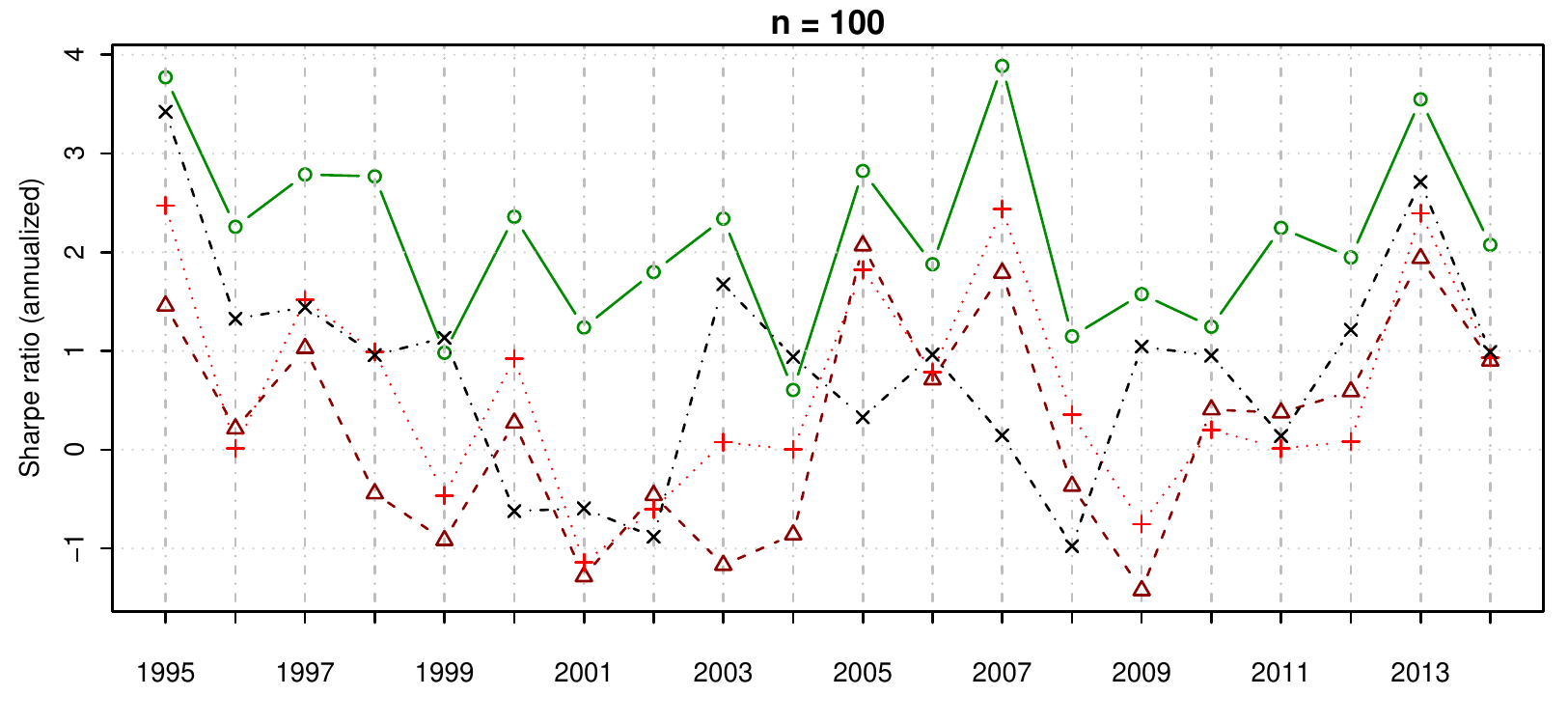}
\par\end{centering}

\begin{centering}
\includegraphics[width=5in]{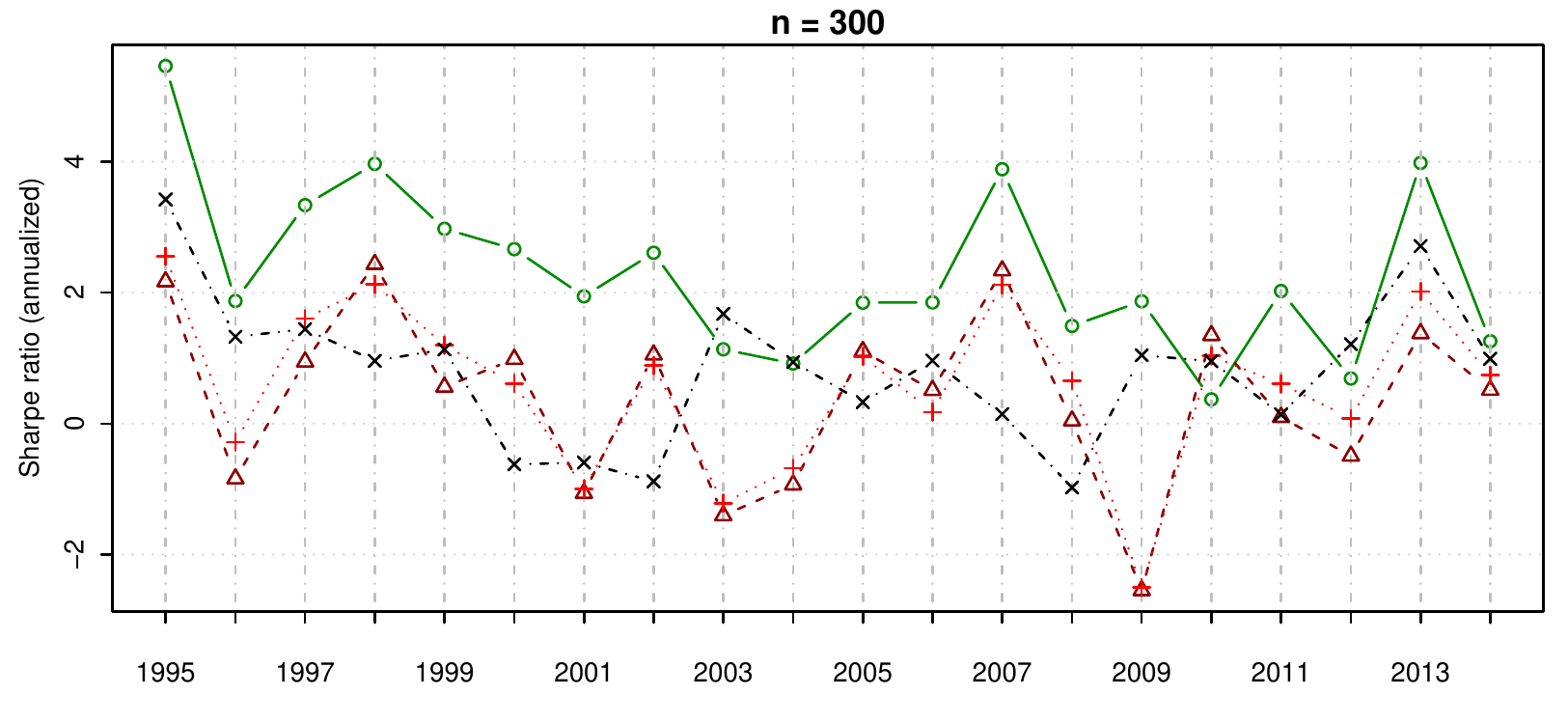}
\par\end{centering}

\begin{centering}
\includegraphics[width=5in]{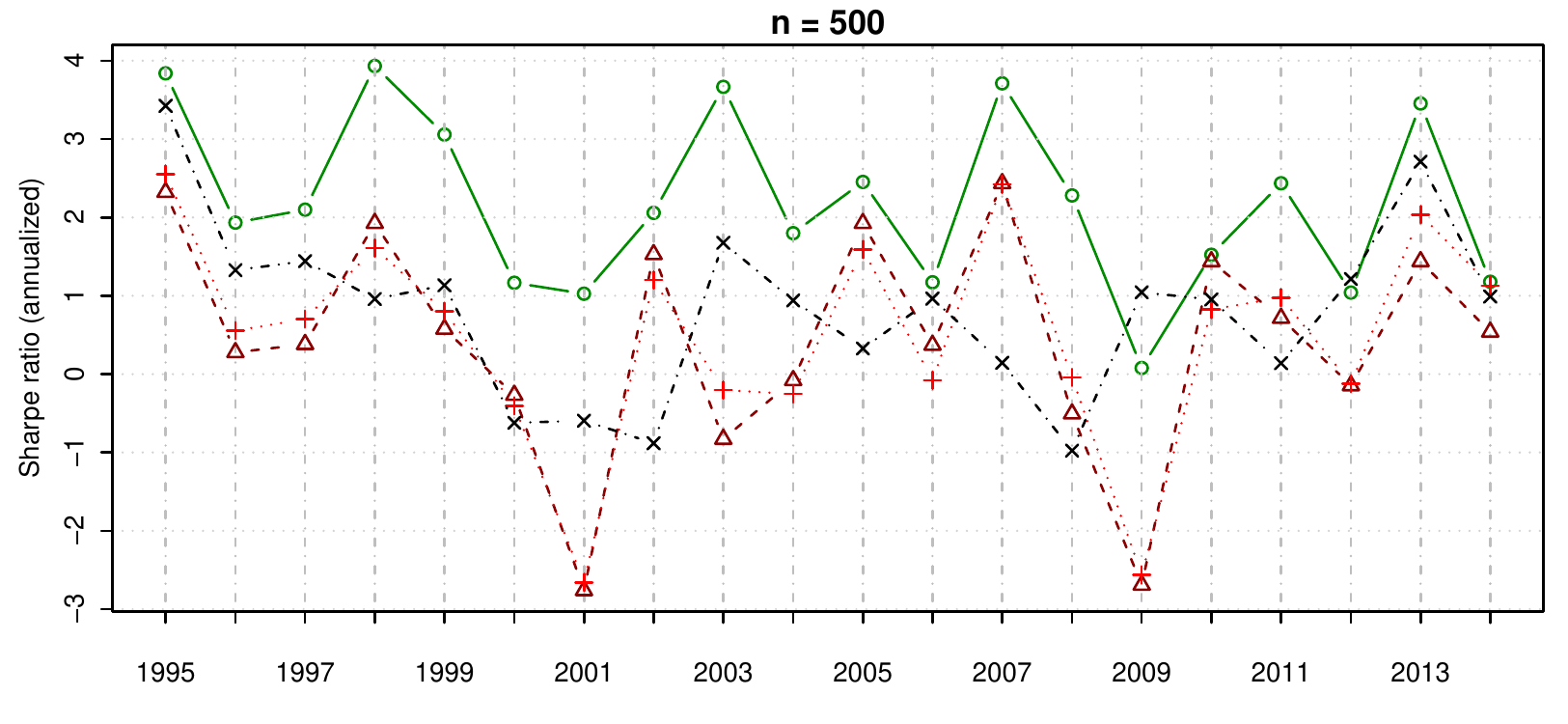}
\par\end{centering}

\begin{centering}
\includegraphics[width=5in]{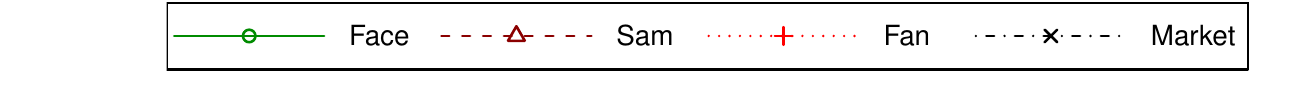}
\par\end{centering}

\begin{singlespace}
\begin{center}
\begin{minipage}{30pc}
\vspace{0.3cm}
{\it This figure shows the performance of the four trading strategies (Face, Sam, Fan and Market) in terms of the annualized
Sharpe ratio, using different sample sizes $n=100$, $n=300$ and $n=500$.}
\end{minipage}
\end{center}
\end{singlespace}

\end{figure}
\par\end{center}
\emph{}

\begin{center}
\begin{figure}[H]
\caption{
\label{fig4}
{\bf Trading strategies during the financial crisis}}

\begin{centering}
\includegraphics[width=5in]{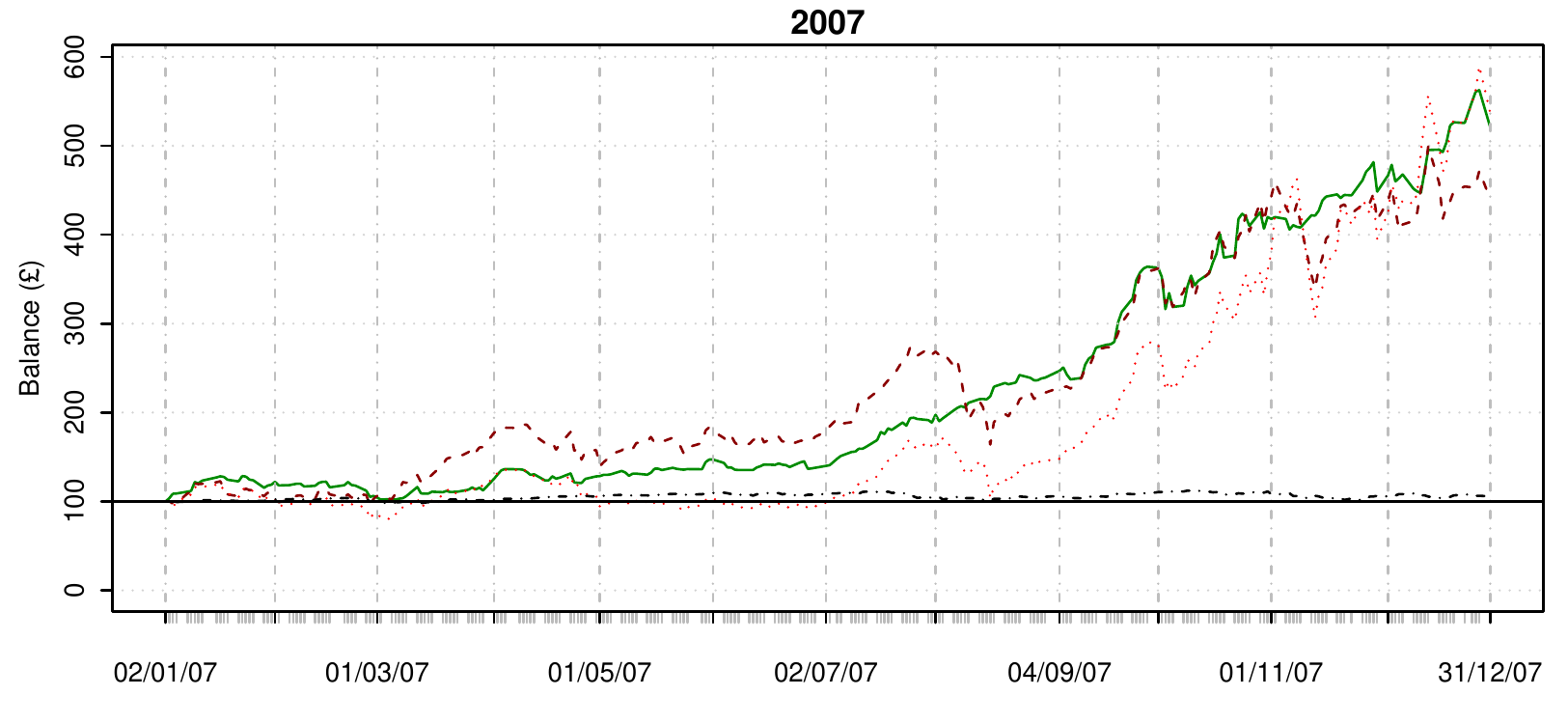}
\par\end{centering}

\begin{centering}
\includegraphics[width=5in]{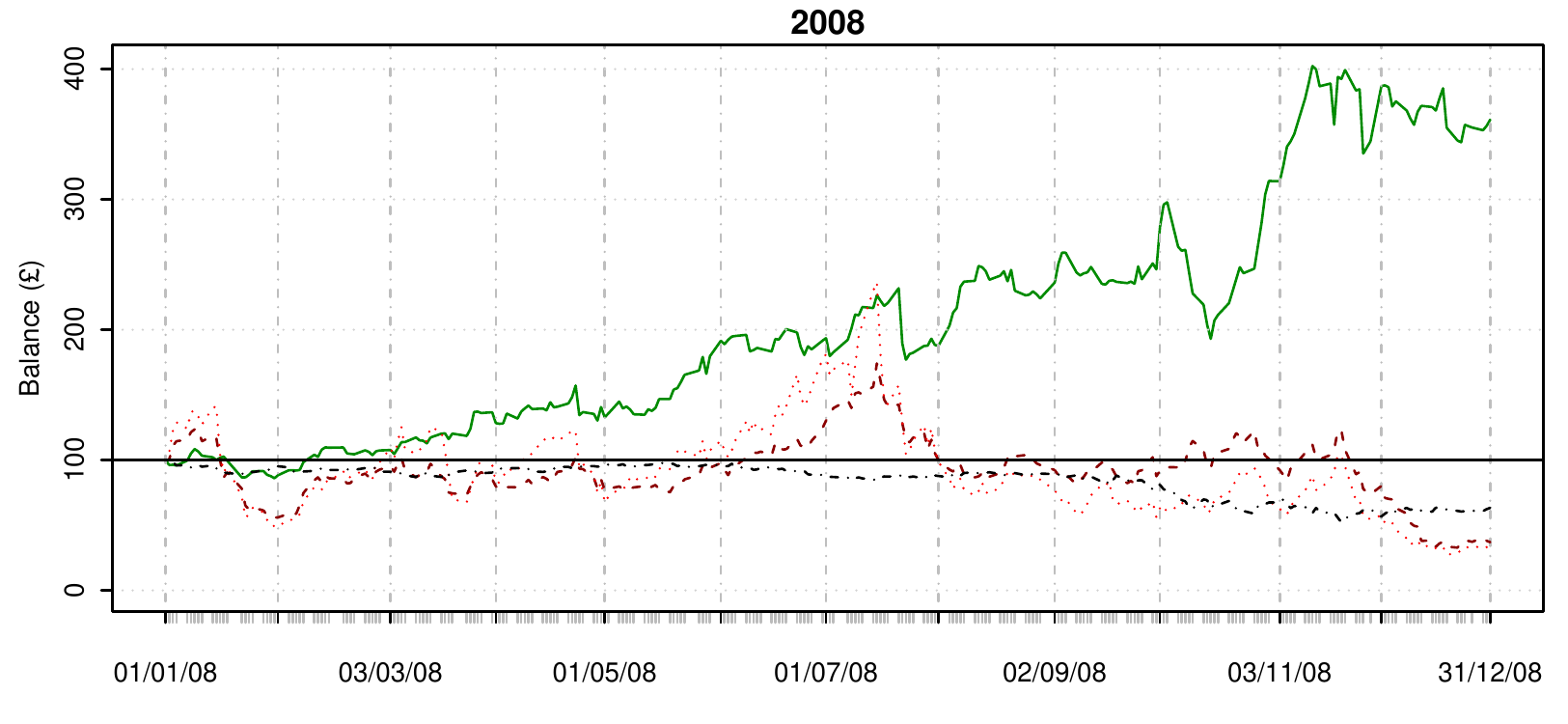}
\par\end{centering}

\begin{centering}
\includegraphics[width=5in]{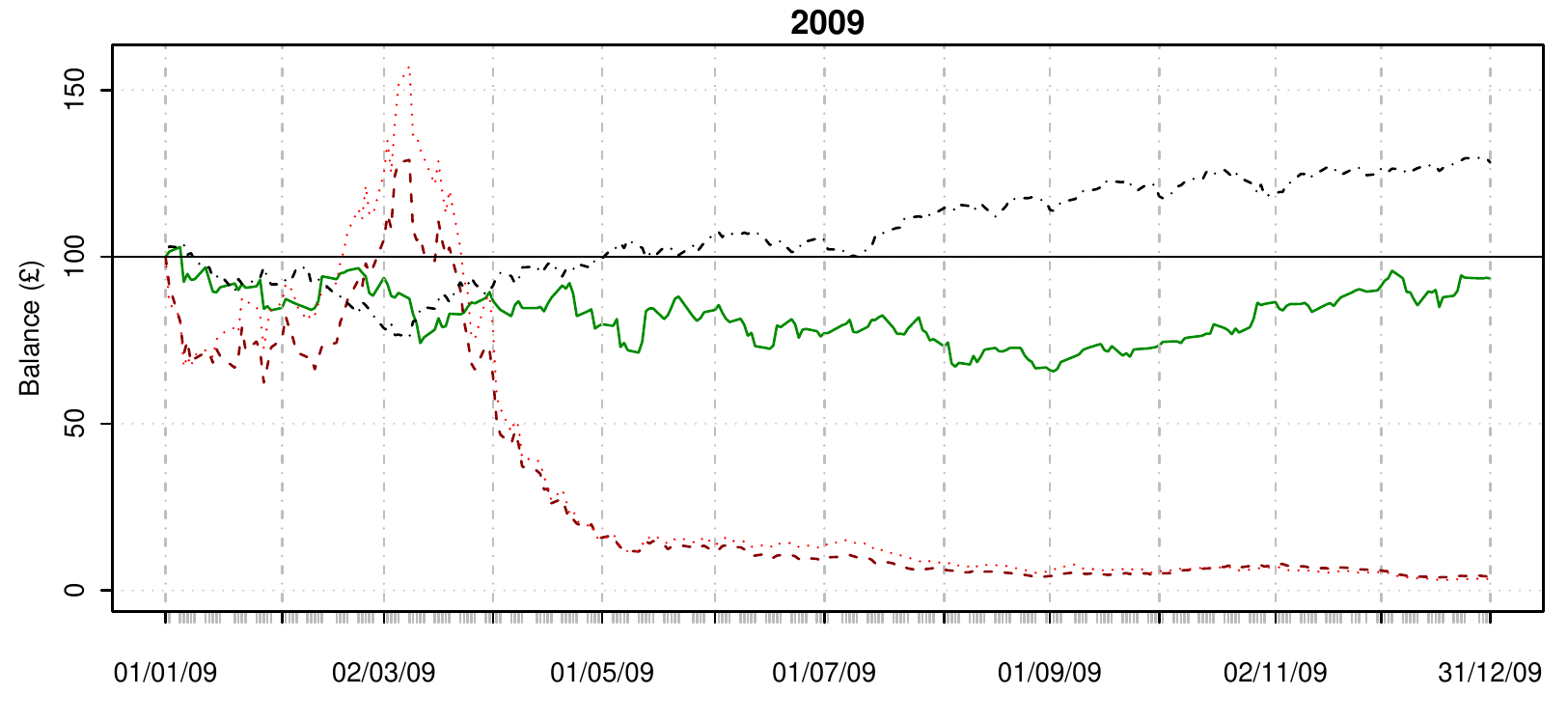}
\par\end{centering}
\centering{}\includegraphics[width=5in]{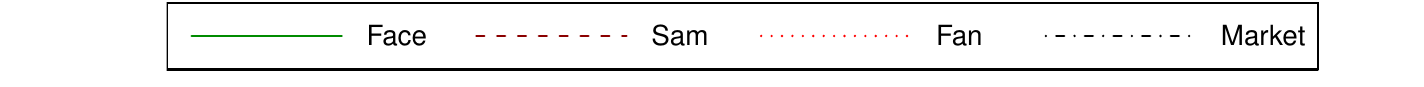}

\vspace{0.3cm}
\begin{singlespace}
\begin{center}
\begin{minipage}{30pc}
{\it This figure shows the performance of the four trading strategies (Face, Sam, Fan and Market) using $n=500$ during 2007, 2008 and 2009 in terms of the end of day balances, assuming an initial balance of 100 pounds at the start of each year.}
\end{minipage}
\end{center}
\end{singlespace}

\end{figure}
\par\end{center}

\begingroup
\renewcommand*{\arraystretch}{0.7}
\begin{center}
\begin{table}[H]
\footnotesize
\begin{centering}
\caption{
\label{tab6}
{\bf Comparison of Balances of Trading Strategies}}
\par\end{centering}
\vspace{2mm}

\begin{centering}
\begin{tabular}{|c|c|ccc|ccc|ccc|}
\noalign{\vskip-0.45cm}
\multicolumn{1}{c}{} & \multicolumn{1}{c}{} &  &  & \multicolumn{1}{c}{} &  &  & \multicolumn{1}{c}{} &  &  & \multicolumn{1}{c}{}\tabularnewline
\noalign{\vskip-0.3cm}
\multicolumn{1}{c}{} & \multicolumn{1}{c}{} &  &  & \multicolumn{1}{c}{} &  &  & \multicolumn{1}{c}{} &  &  & \multicolumn{1}{c}{}\tabularnewline
\hline
\multirow{2}{*}{Year} & \multirow{2}{*}{Market} & \multicolumn{3}{c|}{$n=100$} & \multicolumn{3}{c|}{$n=300$} & \multicolumn{3}{c|}{$n=500$}\tabularnewline
 &  & Face & Sam & Fan & Face & Sam & Fan & Face & Sam & Fan\tabularnewline
\hline
1995 & 137 & 224 & 164 & 216 & 541 & 277 & 347 & 423 & 380 & 466\tabularnewline
1996 & 121 & 159 & 101 & 96 & 184 & 56 & 72 & 212 & 95 & 115\tabularnewline
1997 & 131 & 179 & 138 & 155 & 303 & 146 & 207 & 230 & 98 & 127\tabularnewline
1998 & 124 & 178 & 79 & 134 & 317 & 330 & 299 & 442 & 340 & 273\tabularnewline
1999 & 126 & 121 & 61 & 78 & 260 & 117 & 175 & 329 & 116 & 135\tabularnewline
2000 & 88 & 176 & 102 & 133 & 253 & 155 & 120 & 160 & 54 & 42\tabularnewline
2001 & 89 & 129 & 53 & 60 & 167 & 49 & 49 & 140 & 10 & 6\tabularnewline
2002 & 79 & 164 & 73 & 69 & 222 & 150 & 142 & 196 & 212 & 176\tabularnewline
2003 & 132 & 161 & 57 & 97 & 134 & 40 & 45 & 271 & 53 & 75\tabularnewline
2004 & 112 & 112 & 67 & 95 & 132 & 55 & 56 & 180 & 75 & 63\tabularnewline
2005 & 106 & 179 & 194 & 166 & 184 & 157 & 151 & 265 & 295 & 239\tabularnewline
2006 & 115 & 149 & 119 & 121 & 184 & 114 & 95 & 150 & 103 & 76\tabularnewline
2007 & 106 & 233 & 185 & 231 & 376 & 305 & 321 & 521 & 440 & 537\tabularnewline
2008 & 63 & 143 & 73 & 104 & 203 & 79 & 114 & 361 & 37 & 32\tabularnewline
2009 & 128 & 147 & 48 & 66 & 188 & 9 & 5 & 93 & 4 & 3\tabularnewline
2010 & 117 & 129 & 109 & 100 & 107 & 169 & 148 & 152 & 220 & 140\tabularnewline
2011 & 100 & 177 & 107 & 93 & 192 & 88 & 120 & 283 & 127 & 154\tabularnewline
2012 & 116 & 158 & 117 & 96 & 122 & 60 & 83 & 144 & 71 & 68\tabularnewline
2013 & 135 & 232 & 200 & 226 & 412 & 180 & 275 & 389 & 225 & 363\tabularnewline
2014 & 112 & 158 & 133 & 134 & 152 & 114 & 131 & 162 & 114 & 178\tabularnewline
\hline
\noalign{\vskip-0.3cm}
\multicolumn{1}{c}{} & \multicolumn{1}{c}{} &  &  & \multicolumn{1}{c}{} &  &  & \multicolumn{1}{c}{} &  &  & \multicolumn{1}{c}{}\tabularnewline
\noalign{\vskip-0.45cm}
\multicolumn{1}{c}{} & \multicolumn{1}{c}{} &  &  & \multicolumn{1}{c}{} &  &  & \multicolumn{1}{c}{} &  &  & \multicolumn{1}{c}{}\tabularnewline
\end{tabular}
\par\end{centering}
\vspace{2mm}
\begin{singlespace}
\begin{center}
\begin{minipage}{30pc}
{\it In this table,  
the first two columns show the 
year and the balance on the final trading day when investing in
the market portfolio. 
The balances on the final trading day for Face, Sam and Fan are grouped
according to $n=100$ (columns 3-5), $n=300$  (columns 6-8)
and $n=500$ (columns 9-11).}
\end{minipage}
\end{center}
\end{singlespace}
\end{table}
\par\end{center}
\endgroup


\setcounter{equation}{0}
\renewcommand{\theequation}{A.\arabic{equation}}

 \begin{center}
 {\bf \large APPENDIX}
 \end{center}

 \begin{center}
 {\bf \large Appendix A: Regularity conditions}
 \end{center}

We state the following assumptions.
\begin{description}
\item{Assumption A1.} (i) $\{\bX_t\}_{t\ge 1}$ is stationary and ergodic;
(ii) $\{\ep_t\}_{t \ge 1}$ and $\{\bX_t\}_{t \ge 1}$ are independent; (iii) $\bX_t'$s are bounded with support $\mathcal{X}$, that is, $\sup_{t \ge 1}\|\bX_t\|_\infty \le L, a.s$.
\end{description}

Let $\bbP(A)$ be the probability of a measurable set $A$ and $\bbE(X)$ 
be the expectation of a random variable $X$.  The following strong mixing 
condition (A2) aims at conducting asymptotic properties of the index estimator 
and local linear estimators of nonparametric functions.  Let $\mathcal{F}_{-\infty}^0$ and $\mathcal{F}_{k}^{\infty}$ be the $\sigma-$algebras generated 
by $\{\bX_{t}, t \le 0\}$ and $\{\bX_{t}, t \ge T\}$, respectively 
and define the $\alpha-$mixing coefficient
$$
\alpha(k) = \sup_{A \in \mathcal{F}_{-\infty}^0, B \in \mathcal{F}_{k}^{\infty}} \left | \bbP(A)\bbP(B) - \bbP(AB)\right|.
$$
\begin{description}
\item{Assumption A2.} There exist positive constants $c$ and $0 < \rho < 1$ such that for all $k = 1, 2,\cdots,$
$$
\alpha(k) \le c \rho^{-k}.
$$
\item{Assumption A3.} (i) The kernel function $K(z)$ is a symmetric density function which is bounded with a bounded support and satisfies the Lipschitz condition; (ii) The  density function $f_{\bb}(z)$ of $\bX^{\T} \bb$ is twice differentiable and bounded away from zero on $\{ z = \bx^{\T}\bb; \bx \in \mathcal{X}, \|\bb - \bbeta\|_2 \le c_0\}$ with $0<c_0 <1$; (iii) The density function $f(\bx)$ of $\bX_t$ is bounded away from zero and twice differentiable in $\mathcal{X}$ and the joint densities of $\bX_1$ and $\bX_k$ for all $k \ge 2$ are bounded.
\item{Assumption A4.} $\bg(z)$ and $\bPhi(z)$ have continuous third derivatives in $\mathcal{Z} = \{z: z= \bx^{\T}\bbeta, \bx \in \mathcal{X}\}.$ 
\item{Assumption A5.}  $\|\bV_p - \bV\| = o(1)$, as $p_n \to \infty$, for some $q \times q$ symmetric positive definite $\bV$ such that $\lambda_{\min}(\bV)$ is bounded away from zero.
\end{description}

For the error process $\{\ep_{t}, t \ge 1\}$,  the following assumptions are stated. Denote the true value $\btheta_{\ell} = (\alpha_{\ell,0},\cdots,\alpha_{\ell,m},\gamma_{\ell,1},\cdots,\gamma_{\ell,s})^{\T}$ for $\ell = 1,\cdots,p_n$.
\begin{description}
\item{Assumption B1.} For each $\ell= 1\cdots,p_n$, $\{(\epsilon_{\ell,t},\sigma_{\ell,t}^2), t= 0, \pm 1, \pm 2,\cdots\}$ is a strictly stationary \GARCH$(m,s)$ process with $\sup_{1\le \ell \le p_n}\bbE\sigma_{\ell,1}^{2d} < \infty$ with $d > 4$.
\item{Assumption B2.} Let $\eta_{\ell,t} = \sigma_{\ell,t}^{-1}\epsilon_{\ell,t}$ for each $t$ and $\ell$. Then, for each $\ell = 1,\cdots,p_n,$ the innovations $\eta_{\ell,t}$'s are $i.i.d.$ and absolutely continuous with Lebesgue density being strictly positive in a neighbourhood of zero. Furthermore, $\bbE \eta_{\ell,1}= 0$, $\bbE \eta_{\ell,1}^2 = 1$ and $\sup_{\ell \le p_n}\bbE (\eta_{\ell,1}^{2d}) < \infty$ with $d$ defined in Assumption (B1).

\item{Assumption B3.} For each $\ell=1,\cdots,p_n$, the true value $\btheta_{\ell,0}$ is an interior point of the compact set $ \bLambda$ and $\bLambda \subset (c,+ \infty) \times (c,+\infty)^{m +s} $ for a constant $c >0$.

\item{Assumption B4.} Let $\mathcal{A}_{\ell,\theta}(z) = \sum_{i=1}^m \alpha_{\ell,i} z^i$ and
$\mathcal{B}_{\ell,\theta}(z) = 1- \sum_{i=1}^s \gamma_{\ell,i} z^i$ for $\ell = 1,\cdots,p_n$. If $s>0$, $\mathcal{A}_{\ell,\theta_{\ell,0}}(z)$ and $\mathcal{B}_{\ell,\theta_{\ell,0}}(z)$ have no common roots, $\mathcal{A}_{\ell,\theta_{\ell,0}}(1) \neq 0$, and $\alpha_{\ell,0m} + \gamma_{\ell,0s} \neq 0$.
\end{description}


For the bandwidths $h$, $h_1$, $h_2$ and the dimension $p_n$, we require the following assumptions.
\begin{description}
\item{Assumption C1.} (i) The bandwidth $h$ and $h_1$ satisfy $h = O(n^{-\tau})$ and $h_1 = O(n^{-\tau_1})$, respectively, with $ 1/6 < \tau,\tau_1 < 1/4$.
\item{Assumption C2.} The bandwidth $h_2$ satisfies $h_2 = O(n^{-\tau_2})$ with $ 1/(2q+4) < \tau_2 < 1/(2q+2)$.
\item{Assumption C3.} The dimension $p_n$ satisfies $p_n \le C n^{d/2  - 2 - 2\varepsilon}$ for  some constants $C>0$ and $0 < 2\varepsilon < d/2 -  2$.
\end{description}

Our aim is to estimate $\cov(Y_{t}|\mathcal{F}_{t-1})$. Fan, Fan and Lv (2008) and Fan, Liao and Mincheva (2013) showed that by incorporating the factor structure into the covariance matrix, the resulting estimator has a better convergence rate than the usual sample covariance matrix under the norm $\|\cdot\|_{\bSigma}$. To prove the convergence rate of $\wh{\cov}(Y_t|\mathcal{F}_{t-1}) - \cov(Y_{t}|\mathcal{F}_{t-1})$ under the norm $\|\cdot\|_{\bSigma}$, we impose the following assumption:
\begin{description}
\item{Assumption C4.}
For each $\bx \in \mathcal{X}$, $\|p_n^{-1}\{\bPhi(\bx^{\T}\bbeta)\}^{\T}\bPhi(\bx^{\T}\bbeta) - \bV_2\| = o(1)$, as $p_n \to \infty$ for some $q \times q$ symmetric positive definite $\bV_2$ such that $\lambda_{\min}(\bV_2)$ is bounded away from zero.
\end{description}

The assumptions are regular. The strong mixing condition in the Assumption (A2) can be relaxed as $\alpha(k) \le c k^{-\beta}$ with a large constant $\beta$. Assumption (B1) and (B2) guarantee the existence of the $2d-$th moment of $\epsilon_{\ell,1}$. For simplicity, we do not impose the conditions that ensure the finiteness of the $d-$th moment of $\sigma_{\ell,1}^2$. For more details, see Lindner (2009). Assumption (C4) requires that the factors should be pervasive, that is, impact every individual time series. It was also imposed in Fan, Fan and Lv (2008) and Fan, Liao and Mincheva (2011). 

 \begin{center}
 {\bf \large Appendix B: Proof of Theorem 1 (I)-(III)}
 \end{center}

For ease of presentation, we give some notation. Define
$$
\delta_{\bbb} = \|\bb - \bbeta\|, \delta_{1n} =  \left(\frac{\log (n)}{nh}\right)^{1/2}, \delta_{2n} =  \left(\frac{\log (n)}{n}\right)^{1/2}, \delta_{3n} =  \left(\frac{\log (n)}{nh_1}\right)^{1/2}
$$
and $\tilde{\delta}_n = h^3 + h^2\delta_{1n} + \delta_{1n}^2.
$
Define $\bTheta$ to be a compact set $\{\bb: \|\bb - \bbeta\| \le c_0, \|\bb\| = 1\}$ with a small $c_0 >0$.
For a random sequence $a_n$, $a_n = \bar{O}_{a.s.} (b_n)$ for some sequence $b_n$ means that 
$
P\left\{\|a_n\| > C b_n\right\} = O({n^{-(1+ \varepsilon)}}),
$
where $\varepsilon$ is defined in Assumption (C3).

To prove Theorem 1, the following lemma is useful. 

{\bf Lemma B.1.} Assume that Conditions (A1)-(A3) and (C3) in Appendix A hold and for some $d>4$, 
$$
\sup_{1 \le \ell \le p_n}E |\epsilon_{\ell,t}|^{2d} < \infty,
$$
where $d$ is defined in (C3). Then there exists a constant $C>0$ such that
\begin{eqnarray*}
P\left\{\sup_{1 \le \ell \le p_n}\sup_{(\bb,\bx) \in (\bbTheta,\mathcal{X})} \left|\frac{1}{n} \sum_{t = 1}^n K_{h}(\bX_i^{\T} \bb - \bx^{\T} \bb) \epsilon_{\ell,t}\right| > C \delta_{1n} \right \} \le O\left(\frac{1}{n^{1+\varepsilon}}\right).
\end{eqnarray*}
The proof of Lemma B.1 can be followed from the proof of Lemma 6.1 in Fan and Yao (2003). Of course, some constants involved in the proof need to be modified. For instance, we instead use $B_n= (nh)^{1/2} (\log(n))^{-2}$.

Denote $\bY
= \left(\bY_{2}, \ \cdots, \ \bY_{n}\right),$ $W_{h}(z;\bb)
=
\diag\left\{
K_{h}(\bX_1^{\T} \bb  - z),
\ \cdots, \
K_{h}(\bX_{n-1}^{\T}\bb - z)
\right\}
$ and
$$
\tbX(z;\bb) = \left (\begin{array}{c}
\tbX_{2}^{\T}(z;\bb) \\
\vdots\\
\tbX_{n}^{\T}(z;\bb)
\end{array}
\right)
=
\left(
\begin{array}{cccc}
1 & \bX_2^{\T}  & \bX_1^{\T}\bb - z &
(\bX_1^{\T} \bb - z)\bX_2^{\T}
\\
\vdots & \vdots & \vdots & \vdots
\\
1 & \bX_n^{\T}  & \bX_{n-1}^{\T}\bb - z &
(\bX_{n-1}^{\T} \bb - z)\bX_n^{\T}
\end{array}
\right),
$$
Let $H = \diag(\1_{1 \times (q+1)}, h \1_{1 \times (q+1)})$ and denote
$
\wh{\Omega}_{h}(z;\bb) = H^{-1} \{\tbX(z;\bb)\}^{\T} W_{h}(z;\bb) \tbX(z;\bb)H^{-1}.
$
Denote $\mu_2 = \int u^2 K(u)du$, $\bmu_{\bbb}(z) = \bbE(\bX| \bX^{\T}\bb = z)$ and, for $ \ell = 1,\cdots,p_n$, 
$\tilde{\ep}_{\ell} = (\epsilon_{\ell,2}, \cdots,\epsilon_{\ell,n})^{\T}$,
\begin{eqnarray*}
\wh{\bGamma}_{\ell}(z;\bb) &=& H^{-1}\left \{\wh{\Omega}_{h}(z;\bb)\right \}^{-1} H^{-1}\tbX^{\T}(z;\bb) W_{h}(z;\bb) \by_{\ell},
\bGamma_{\ell}(z) = (g_{\ell}(z),(\bphi_{\ell}(z))^{\T}, \dot{g}_{\ell}(z), (\dot{\bphi}_{\ell}(z))^{\T})^{\T},\\
\bGamma'_{\ell}(z) &=& (\dot{g}_{\ell}(z),(\dot{\bphi}_{\ell}(z))^{\T}, \ddot{g}_{\ell}(z), (\ddot{\bphi}_{\ell}(z))^{\T})^{\T},
\bGamma''_{\ell}(z) = (\ddot{g}_{\ell}(z),(\ddot{\bphi}_{\ell}(z))^{\T}, \0_{1 \times (q+1)})^{\T}.
\end{eqnarray*}

The following lemma gives the asymptotic representation of $\wh{\bGamma}_{\ell}(z)$.

{\bf Lemma B.2.} Suppose that Assumption (A1)-(A4) in Appendix A hold. Then we have that
\begin{eqnarray*}
H\wh{\bGamma}_{\ell}(z;\bb)
&=&
H\bGamma_{\ell}(z) + \big\{ \wh{\Omega}_{h}(z;\bb)\big\}^{-1}H^{-1}\tbX^{\T}(z;\bb) W_{h}(z;\bb ) \tilde{\ep}_{\ell} + H \bGamma'_{\ell}(z) \big(\bmu_{\bbb}(z)\big)^{\T}(\bbeta - \bb) \\
&& +
\frac{1}{2}\mu_2 h^2 H \bGamma''_{\ell}(z) + \bar{O}_{a.s.}\big(h\delta_{\bbb} + \delta_{1n} \delta_{\bbb} + \delta_{\bbb}^2 + \tilde{\delta}_n\big).
\end{eqnarray*}

{\bf Proof of Lemma B.2.} For $i = 2,\cdots,n$, denote $z_i = \bX_{i-1}^{\T}\bbeta$ and $z_{\bbb,i} = \bX_{i-1}^{\T}\bb$. Using a Taylor's expansion, we obtain that
$$
y_{\ell,i} = g_{\ell}(z_i) + \bphi_{\ell}(z_i) \bX_i + \epsilon_{\ell,i} =  \tbX_{i}^{\T}(z;\bb) \bGamma_{\ell}(z)+ \epsilon_{\ell,i} + r_{\ell,\bbb,i}^{(1)} + r_{\ell,\bbb,i}^{(2)} +r_{\ell,\bbb,i}^{(3)} + r_{\ell,\bbb,i}^{(4)},
$$
where
$ r_{\ell,\bbb,i}^{(1)} = \tbX_{\bbb,i}^{\T}(z)\bGamma'_{\ell}(z) (z_{i} - z_{\bbb,i}),
 r_{\ell,\bbb,i}^{(2)} = 2^{-1} \tbX_{\bbb,i}^{\T}(z)\bGamma''_{\ell}(z) (z_{\bbb,i} - z)^2,$ \\
 $
 r_{\ell,\bbb,i}^{(3)} = 2^{-1} \tbX_{\bbb,i}^{\T}(z)\bGamma''_{\ell}(z) (z_i - z_{\bbb,i} )^2,
 r_{\ell,\bbb,i}^{(4)} = O(|z_i -z|^3).
$

For $k=1,\cdots, 4$, denote $\br_{\ell,\bbb}^{(k)} = (\br_{\ell,\bbb,2}^{(k)}, \cdots, \br_{\ell,\bbb,n}^{(k)})^{\T}$. Then
\begin{eqnarray*}
H\wh{\bGamma}_{\ell}(z;\bb)-  H\bGamma_{\ell}(z) = \left \{\wh{\Omega}_{h}(z;\bb)\right \}^{-1} H^{-1}\tbX^{\T}(z;\bb) W_{h}(z;\bb) \left(\tilde{\ep}_{\ell} + \br_{\ell,\bbb}^{(1)} + \br_{\ell,\bbb}^{(2)} + \br_{\ell,\bbb}^{(3)} + \br_{\ell,\bbb}^{(4)}\right).
\end{eqnarray*}

{ (I).} Consider the term $\wh{\Omega}_{h}(z;\bb)$. Following the proof of Theorem 5.3 in Fan and Yao (2003), we have that there exists a large $C>0$ such that
\begin{eqnarray*}
\bbP\left \{ \sup_{(\bbb,z) \in \bbTheta \times \mathcal{Z}} \left\|{1 \over n}\left(\wh{\Omega}_{h}(z;\bb) - \bbE\left\{\wh{\Omega}_{h}(z;\bb)\right\}\right)\right\|_F > C \delta_{1n}\right \} \le O\left({1\over n^2}\right).
\end{eqnarray*}
Let $\Omega(z;\bb) = \lim_{n \to \infty} n^{-1}\bbE\left\{\wh{\Omega}_{h}(z;\bb)\right\}$. Note that
$
n^{-1}\bbE\wh{\Omega}_{h}(z;\bb) = \Omega(z;\bb) + O(h)
$
and $\Omega(z;\bb)$ is positive definite.
Therefore, $\wh{\Omega}_{h}(z;\bb)$ is positive definite almost surely and
$$
n^{-1}\wh{\Omega}_{h}(z;\bb) = \Omega(z;\bb) + \bar{O}_{a.s.}(h + \delta_{1n}).
$$

{(II).} Consider the term $H^{-1} \tbX^{\T}(z;\bb) W_{h}(z;\bb)\br_{\bbb,\ell}^{(k)}$ $(k = 1,\cdots,4)$.  By specific matrix calculations,  we can show that
\begin{eqnarray*}
H^{-1} \tbX^{\T}(z;\bb) W_{h}(z;\bb)\br_{\bbb,\ell}^{(1)} &=& \Omega(z;\bb)H \bGamma'_{\ell}(z) \big(\bmu_{\bbb}(z)\big)^{\T}(\bbeta - \bb) +  \bar{O}_{a.s.}\Big ( h\delta_{\bbb} + \delta_{1n} \delta_{\bbb} \Big),\\
H^{-1} \tbX^{\T}(z;\bb) W_{h}(z;\bb)\br_{\bbb,\ell}^{(2)} &=& \frac{1}{2}\mu^2 h^2 \Omega_(z;\bb)H \bGamma''_{\ell}(z) +   \bar{O}_{a.s.}\Big (h^3 + h^2\delta_{1n} \Big ),\\
H^{-1} \tbX^{\T}(z;\bb) W_{h}(z;\bb)\br_{\bbb,\ell}^{(3)} &=&
\bar{O}_{a.s.} \big(\delta_{\bbb}^2\big), \ \
H^{-1} \tbX^{\T}(z;\bb) W_{h}(z;\bb)\br_{\bbb,\ell}^{(4)} = \bar{O}_{a.s.} \big(\delta_{\bbb}^3 + h^3 + h^2 \delta_{\bbb} + h \delta_{\bbb}^2\big).
\end{eqnarray*}

Combining (I) and (II), we obtain that
\begin{eqnarray*}
H\wh{\bGamma}_{\ell}(z;\bb)
&=&  H\bGamma_{\ell}(z) + \big\{ \wh{\Omega}_{h}(z;\bb)\big\}^{-1}H^{-1}\tbX^{\T}(z;\bb) W_{h}(z;\bb) \tilde{\ep}_{\ell} + H \bGamma'_{\ell}(z) \big(\bmu_{\bbb}(z)\big)^{\T}(\bbeta - \bb) \\
&& +
\frac{1}{2}\mu^2 h^2 H \bGamma''_{\ell}(z) + \bar{O}_{a.s.}\big(h\delta_{\bbb} + \delta_{1n} \delta_{\bbb} + \delta_{\bbb}^2 + \tilde{\delta}_n\big).
\end{eqnarray*}
This completes the proof.

The following lemma, Lemma B.3, gives the asymptotic relationship between $\wh{\bbeta}_{m+1}$ and $\wh{\bbeta}_{m}$, where $\wh{\bbeta}_{m}$ is the $m$th step estimator based on our procedure in Section 2.

Without loss of generality, we consider  $m = 1$.
For each $i,j= 1,\cdots,n-1$, define
$$
X_{ij} = X_i - X_j, w_{ij}(\bb) = h^{-1}K\left\{X_{ij}^{\T} \bb/{h}\right\}.
$$
Given $\wh{\bbbeta}_1$, for $j = 1,\cdots, n-1$, denote $\hat{z}_j = \bX_j^{\T}\wh{\bbbeta}_{1}$ and
 $$
\wh{\bGamma}_{j} = (\wh{\bg}_j, \wh{\bxi}_j, \wh{A}_j, \wh{B}_j) = \bY W_{h}(\hat{z}_j;\wh{\bbbeta}_1) \tbX(\hat{z}_j;\wh{\bbbeta}_1) \left\{\tbX^{\T}(\hat{z}_j;{\wh{\bbbeta}_1})W_{h}(\hat{z}_j;\wh{\bbbeta}_1) \tbX(\hat{z}_j;{\wh{\bbbeta}_1})\right\}^{-1}.
$$
and
\begin{eqnarray*}
\wh{\bV}_n
&=&
\frac{1}{n^2p_n}\sum_{i,j=1}^{n-1}\bX_{ij}\bX_{ij}^{\T}
\|\wh{\bxi}_j + \wh{B}_j \bX_{i+1}\|^2 w_{ij}(\wh{\bbeta}_1),\\
\wh{\bU}_n &=& \frac{1}{n^2p_n}\sum_{i,j=1}^{n-1}\bX_{ij}\Big (\wh{\bxi}_j + \wh{B}_j\bX_{i+1}\Big )^{\T}\Big \{\bY_{i+1}- \wh{\bGamma}_{j} \widetilde{\bX}(\hat{z}_j;\wh{\bbeta}_1)\Big \} w_{ij}(\wh{\bbeta}_1),\\
\wh{\bbeta}_2 &=& \wh{\bbeta}_1 + \wh{\bV}^{-1}_n \wh{\bU}_n.
\end{eqnarray*}

{\bf Lemma B.3. } Suppose that Conditions (A1)-(A4), (B1)-(B4), (C1) and (C3) in Appendix A hold. Then, we have
\begin{eqnarray}
\wh{\bbeta}_2 - \bbeta = \frac{1}{2}\left(\wh{\bbeta}_1 - \bbeta\right ) + \frac{1}{2} \bV_p^{-1}\bU_n + \bR_n,
\end{eqnarray}
where $\bR_n = \bar{O}_{a.s.}\left(h \delta_{2n} + h^{-1}\delta_{2n}^2 + \tilde{\delta}_n + h^{-1}\delta_{2n} \delta_{\wh{\bbbeta}_1}+h\delta_{\wh{\bbbeta}_1} + h^{-1}\delta^2_{\wh{\bbbeta}_1}\right).$

{\bf Proof of Lemma B.3.} First, consider the term $\bU_n$. For $i,j =1,\cdots,n-1,$ denote
\begin{eqnarray*}
\be_{ij,1} &=& \bg'(\bX_j^{\T}\bbeta) + \bPhi'(\bX_j^{\T}\bbeta )\bX_{i+1}, 
\be_{ij,2} = \wh{\bxi}_j + \wh{\bB}_j\bX_{i+1} - \be_{ij,1},\\
\be_{i,3} &=& \bg(\hat{z}_i) + \bPhi(\hat{z}_i)\bX_{i+1} + (\bg'(\hat{z}_i) + \bPhi'(\hat{z}_i)\bX_{i+1}) (\bmu_{\bbbeta}(\bX_i^{\T}\bbeta))^{\T}(\bbeta - \wh{\bbeta}_1 ),\\
\be_{ij,4} &=& \wh{\bGamma}_{j} \widetilde{\bX}_{i+1}(\hat{z}_j;\wh{\bbeta}_1)- \be_{i,3}.
\end{eqnarray*}

We decompose $\wh{\bU}_n$ as
\begin{eqnarray*}
\wh{\bU}_n &=& \frac{1}{n^2p_n}\sum_{i,j=1}^{n-1}\bX_{ij}\be_{ij,1}^{\T}\big (\bY_{i+1}- \be_{i,3}\big ) w_{ij}(\wh{\bbeta}_1)
- \frac{1}{n^2p_n}\sum_{i,j=1}^{n-1}\bX_{ij}\be_{ij,1}^{\T}\be_{j,4} w_{ij}(\wh{\bbeta}_1)\\
& & + \frac{1}{n^2p_n}\sum_{i,j=1}^{n-1}\bX_{ij}\be_{ij,2}^{\T}\big (\bY_{i+1}- \be_{i,3}\big ) w_{ij}(\wh{\bbeta}_1)
 - \frac{1}{n^2p_n}\sum_{i,j=1}^{n-1}\bX_{ij}\be_{ij,2}^{\T}\be_{ij,4} w_{ij}(\wh{\bbeta}_1) = \sum_{k=1}^4\bU_{nk}.
\end{eqnarray*}

{\bf (a).} Consider the main term $\bU_{n1}$. Note that
$$
\bY_{i+1}- \be_{i,3} = \ep_{i+1} + \left(\bg'(\bX_i^{\T}\bbeta) + \bPhi'(\bX_i^{\T}\bbeta)\bX_{i+1} \right) \big(\bX_i - \bmu_{\bbbeta}(\bX_i^{\T}\bbeta)\big)^{\T}(\bbeta - \wh{\bbeta}_1) + O(\delta_{\wh{\bbbeta}_1}^2).
$$
Analogous to Lemma A.2 of Xia, Tong and Li (2002), it follows that
$$
\frac{1}{n^2p_n}\sum_{i,j=1}^{n-1}\bX_{ij}\be_{ij,1}^{\T} \ep_{i+1} w_{ij}(\wh{\bbeta}_1) 
= \bU_n + \bar{O}_{a.s.} (\delta_{1n}\delta_{\wh{\bbbeta}_1}).
$$
Similarly, we obtain that
\begin{eqnarray*}
\frac{1}{n^2p_n}\sum_{i,j=1}^{n-1}\bX_{ij}\be_{ij,1}^{\T}\left(\bg'(\bX_i^{\T}\bbeta) + \bPhi'(\bX_i^{\T}\bbeta)\bX_{i+1} \right) \big(\bX_i - \bmu_{\bbbeta}(\bX_i^{\T}\bbeta)\big)^{\T}w_{ij}(\wh{\bbeta}_1) = \bV_p + \bar{O}_{a.s.}(\delta_{1n} + \delta_{\wh{\bbbeta}_1}).
\end{eqnarray*}
Hence, we approximate the term $\bU_{n1}$ as 
\begin{eqnarray*}
\bU_{n1} = \bU_{n} + {\bV}_p (\bbeta - \wh{\bbeta}_1) + \bar{O}_{a.s.} \big( \delta_{1n}\delta_{\wh{\bbbeta}_1}  + \delta_{\wh{\bbbeta}_1}^2\big).
\end{eqnarray*}

{\bf (b).} With the help of asymptotic representation of $\wh{\bGamma}_j(z)$ and empirical approximation theories, we can show that
$$
\bU_{nk} = \bar{O}_{a.s.}\big(h\delta_{\wh{\bbbeta}_1} + h^{-1}\delta_{2n} \delta_{\wh{\bbbeta}_1} + h^{-1}\delta_{\wh{\bbbeta}_1}^2 + \tilde{\delta}_n\big), k = 2,3,4.
$$

{\bf (c).} In the similar fashion, we can also show that
$$
\wh{\bV}_n = 2\bV_p + \bar{O}_{a.s.} \left( \delta_{\wh{\bbbeta}_1} + h + h^{-1}\delta_{2n} \right ).
$$
Therefore,
$
\wh{\bbeta}_2 - \wh{\bbeta}_1 = 2^{-1}(\bbeta - \wh{\bbeta}_1) + 2^{-1}{\bV}_p^{-1}{\bU}_n + \bR_n,
$
which means that
\begin{eqnarray}
\wh{\bbeta}_2 - \bbeta = \frac{1}{2}( \wh{\bbeta}_1 - \bbeta) + \frac{1}{2}{\bV}_p^{-1}{\bU}_n + \bR_n.
\end{eqnarray}
This completes the proof.

{\bf Proof of Theorem 1 (I).} First, by Lemma B.3, for the $m-$th step ($m > 1$), we have
\begin{eqnarray}
\label{iteration}
\wh{\bbeta}_{m+1} - \bbeta = \frac{1}{2}( \wh{\bbeta}_m - \bbeta ) + \frac{1}{2} \bV_p^{-1}{\bU}_n + \bR_{n,m},
\end{eqnarray}
where
$\|\bR_{n,m}\| \le M \left(\delta_{\wh{\bbbeta}_m}(h + h^{-1}\delta_{2n} + h^{-1}\delta_{\wh{\bbbeta}_m}) + \tilde{\delta}_{n} + h \delta_{2n} + h^{-1}\delta_{2n}^2 \right) $
a.s. and \\ $\|{\bV}_p^{-1}{\bU}_n\| \le M \delta_{2n}$ a.s., with some large positive constant $M$. Here we take $M > 1$ and $h < 1$ for sufficiently large $n$.
Note that as $n \to \infty$,  the bandwidth $h$ satisfies $h \to 0$, $h^{-1}\delta_{2n} \to 0$, $\tilde{\delta}_nh^{-1} \to 0$ and $h^{-2}\delta_{2n}^2 \to 0$. We can assume that
$$
h + h^{-1}\delta_{2n} \le (8M)^{-1}, M(\tilde{\delta}_n + h \delta_{2n} + h^{-1}\delta_{2n}^2) + M\delta_{2n} \le (32M)^{-1}h.
$$
Then, if $\delta_{\bbbeta_{m}} \le (8M)^{-1}h$, then $\delta_{\bbbeta_{m+1}} \le (8M)^{-1}h$ and
$$
\delta_{\bbbeta_{m+1}} \le \frac{3}{4}\delta_{\bbbeta_{m}} + M(\tilde{\delta}_n + h \delta_{2n} + h^{-1}\delta_{2n}^2) + M\delta_{2n}.
$$
Note that we can choose the initial estimator $\wh{\bbeta}_1$ which satisfies
$\|\delta_{\bbbeta_{1}}\| \le (8M)^{-1}h$ for sufficiently large $n$. Therefore,
$$
\delta_{\bbbeta_{m+1}} \le \left(\frac{3}{4}\right)^{m} \delta_{\bbbeta_{1}}  +
\left\{1 + \frac{3}{4} + \cdots + \left(\frac{3}{4}\right)^m\right\} \left\{M(\tilde{\delta}_n + h \delta_{2n} + h^{-1}\delta_{2n}^2 +\delta_{2n})\right\}.
$$
Taking $m \to \infty$, it follows that the final estimator $\wh{\bbbeta}$ satisfies $\delta_{\bbbeta} = \|\wh{\bbbeta} - \bbbeta\| = \bar{O}_{a.s.}\big(\tilde{\delta}_n + \delta_{2n} + h^{-1}\delta_{2n}^2\big)$ and hence
$
\|\bR_{n,\infty}\| = \bar{O}_{a.s.}\left(h^3 + \delta_{1n}^2\right).
$
It also follows from the expression (\ref{iteration}) that 
$$
P\left \{ \|\wh{\bbeta}  - \bbeta - \bV_p^{-1} \bU_n\| \ge C \left(h^3 + \delta_{1n}^2\right)\right\} \le O\left(\frac{1}{n^{1 + \varepsilon}}\right).
$$
This completes the proof of Theorem 1(I). 

{\bf Proof of Theorem 1 (II) and (III).} Lemma B.2 tells us that, for $ \ell = 1,\cdots,p_n$,
\begin{eqnarray*}
\wh{g}_{\ell}(z) - g_{\ell}(z) =  e^{\T}_1\big\{ \wh{\Omega}_{h_1}(z;\wh{\bbeta})\big\}^{-1}H_1^{-1}\tbX^{\T}(z;\wh{\bbeta}) W_{h_1}(z;\wh{\bbeta}) \tilde{\ep}_{\ell} + \frac{1}{2}\mu_2 h_1^2 \ddot{g}_{\ell}(z) + \bR_n(z),
\end{eqnarray*}
where $e_1 = (1,0,\cdots,0)^{\T}$, $H_1 = \diag(\1_{1\times (q+1)}, h_1\1_{1\times (q+1)})$ and
$$
\bbP \Big\{\sup_{z \in \mathcal{Z}}|\bR_n(z)| > C (h^3 + \delta_{1n}^2 + n^{-1/2}) \Big\} = O\left(\frac{1}{n^{1+\varepsilon}}\right).
$$
for some constant $C>0$.

{\bf (a).} Consider the term $\wh{\Omega}_{h_1}(z;\bb)$. Following the proof of Theorem 5.3 in Fan and Yao (2003), we have that there exists a large $C>0$ such that
\begin{eqnarray*}
\bbP\left \{ \sup_{(\bbb,z) \in \bbTheta \times \mathcal{Z}} \left\|{1 \over n}\left(\wh{\Omega}_{h_1}(z;\bb) - \bbE\left\{\wh{\Omega}_{h_1}(z;\bb)\right\}\right)\right\|_F > C \delta_{3n}\right \} \le O\left({1\over n^2}\right).
\end{eqnarray*}
Let $\Omega(z;\bb) = \lim_{n \to \infty} n^{-1}\bbE\left\{\wh{\Omega}_{h_1}(z;\bb)\right\}$. Note that
$
n^{-1}\bbE\wh{\Omega}_{h_1}(z;\bb) = \Omega(z;\bb) + O(h_1)
$
and $\Omega(z;\bb)$ is positive definite.
Therefore, $\wh{\Omega}_{h_1}(z;\bb)$ is positive definite almost surely and
\begin{eqnarray*}
P\left \{ \sup_{(\bbb,z) \in \bbTheta \times \mathcal{Z}} \left\|{1 \over n}\wh{\Omega}_{h_1}(z;\bb) - {\Omega}(z;\bb)\right\|_2^2 > C (h_1+ \delta_{3n})\right \} \le O\left({1\over n^2}\right).
\end{eqnarray*}

{\bf (b).} By Lemma B.1, we have
\begin{eqnarray*}
P\left( \sup_{1 \le \ell \le p_n }\sup_{(\bbb,z) \in \bbTheta \times \mathcal{Z}}  \|{1 \over n}H_1\tbX^{\T}(z;\bb) W_{h_1}(z;\bb)\tilde{\ep}_{\ell}\|_2 > C \delta_{3n}\right)  \le O\left({1\over n^{1+\varepsilon}}\right).
\end{eqnarray*}
Therefore, combining (a) and (b), there exists a large $C>0$ such that
\begin{eqnarray*}
\bbP \left\{ \sup_{z \in \mathcal{Z} } \left\|\wh{\bg}(z) - \bg(z) \right\|_{\infty} > C (h_1^2 + \delta_{3n})\right \}  \le O\left({1\over n^{1+\varepsilon}}\right).
\end{eqnarray*}
This completes the proof of Theorem 1(II). Theorem 1(III) can be proven analogously.

 \begin{center}
 {\bf \large Appendix C: Proof of Theorem 1 (IV)}
 \end{center}

Before we prove Theorem 1(IV), we first give the convergence rate of the difference between the estimated residual $\wh{\ep}_{t}$ and the true residual $\ep_t$.

{\bf Lemma C.1.} Suppose that Assumptions (A1)-(A5), (B1)-(B4) and (C1) and (C3) in Appendix A hold. Then there exists $C > 0$ and small $\varepsilon >0$ such that
\begin{eqnarray*}
\bbP \left\{\sup_{t \le n}\left\|\wh{\ep}_{t} - \ep_{t}\right\|_{\infty} > C \left( h_1^2 + \delta_{3n}\right) \right\} \le O\left({1\over n^{1+ \varepsilon}} \right).
\end{eqnarray*}

{\bf Proof of Lemma C.1.} For each $t =2, \cdots,n,$
$$
\wh{\ep}_{t} - \ep_{t}  = \wh{\bg}(\bX_{t-1}^{\T}\wh{\bbeta})  - \bg(\bX_{t-1}^{\T}\bbeta) + \left(\wh{\bPhi}(\bX_{t-1}^{\T} \wh{\bbeta}) -  \bPhi(\bX_{t-1}^{\T} \bbeta)\right)\bX_{t}.
$$
Note that
$
 \wh{\bg}(\bX_{t-1}^{\T}\wh{\bbeta})  - \bg(\bX_{t-1}^{\T}\bbeta)
 =  \bg'(\bX_{t-1}^{\T}\bbeta^*) \bX_{t-1}^{\T} (\wh{\bbeta} - \bbeta)
+ \wh{\bg}(\bX_{t-1}^{\T}\wh{\bbeta})  - \bg(\bX_{t-1}^{\T}\wh{\bbeta}),
$
 and
 \begin{eqnarray*}
(\wh{\bPhi}(\bX_{t-1}^{\T}\wh{\bbeta})  - \bPhi(\bX_{t-1}^{\T}\bbeta) ) \bX_{t}
& = &\bPhi'(\bX_{t-1}^{\T}\bbeta^* ) \bX_{t} \bX_{t-1}^{\T} (\wh{\bbeta} - \bbeta) \\
& & + (\wh{\bPhi}(\bX_{t-1}^{\T}\wh{\bbeta})  - \bPhi(\bX_{t-1}^{\T}\wh{\bbeta}))\bX_{t}.
\end{eqnarray*}
Hence, there exists a large constant $C>0$ such that
\begin{eqnarray*}
\|\wh{\ep}_{t} - \ep_{t}\|_{\infty} \le \sup_{z \in \mathcal{Z}} \|\wh{\bg}(z) - \bg(z)\|_{\infty} +
\sup_{z \in \mathcal{Z}} \|\wh{\bPhi}(z) - \bPhi(z)\|_{\infty} + C \|\wh{\bbeta} - \bbeta\|,
\end{eqnarray*}
where $\sup_{z \in \mathcal{Z}} \left(\|\bg'(z)\|_{\infty} + \|\bPhi'(z)\|_{\infty} \right) = O(1)$ is used in the last terms. 
For any $v>0$, we have the following inequality
\begin{eqnarray*}
 \bbP \left \{ \sup_{ 2 \le t \le n} |\wh{\ep}_{t} - \ep_{t}| > 3 v \right \} &\le & \bbP\left \{ \|\wh{\bbeta} - \bbeta_{0}\| > v/C\right\}  + P\left\{\sup_{ z \in \mathcal{Z}} \left\|\wh{\bg}(z)  - \bg(z) \right\|_{\infty} > v \right\} \\
&& + P\left\{\sup_{ z \in \mathcal{Z}} \left\|\wh{\bPhi}(z)  - \bphi(z) \right\|_{\infty} > v \right\}.
\end{eqnarray*}
Take $v = C(h_1^2 + \delta_{3n})$ for a large constant $C>0$. It follows from parts (II) and (III) of Theorem 1 that there exists a constant $C>0$ such that
\begin{eqnarray*}
\bbP\Big\{\sup_{2 \le t \le n}\left\|\wh{\ep}_{t} - \ep_{t}\right\|_{\infty} > C\left(h_1^2 + \delta_{3n}\right)\Big\} \le O\left(\frac{1}{n^{1+ \varepsilon}}\right).
\end{eqnarray*}
This completes the proof of Lemma C.1.

Now we are going to prove Theorem 1(IV). Define the quasi log-likelihood function
$$
\wt{Q}_{\ell,n}(\btheta) = n^{-1} \sum_{t = 1}^n \wt{v}_{\ell,t}(\btheta), \wt{v}_{\ell,t}(\btheta) = \frac{\epsilon_{\ell,t}^2}{\wt \sigma_{\ell,t}^2(\btheta)} + \log \wt{\sigma}_{\ell,t}^2(\btheta),
$$
where $\wt{\sigma}_{\ell,t}^2(\btheta)$ is the solution of
$$
\wt{\sigma}_{\ell,t}^2(\btheta) = \alpha_{\ell,0} + \sum_{i=1}^m \alpha_{\ell,i}\epsilon_{\ell,t-i}^2
+ \sum_{i= 1}^s \gamma_{\ell,i} \wt{\sigma}_{\ell,t-i}^2(\btheta).
$$
For convenience, denote the true value of $\btheta_{\ell}$ by $\btheta_{\ell,0}$. 
First, we consider the consistency of $\wh{\btheta}_{\ell}$. Recall that the observed quasi log likelihood function 
$$
{Q}_{\ell,n}(\btheta) = n^{-1} \sum_{t = 1}^n {v}_{\ell,t}(\btheta), {v}_{\ell,t}(\btheta) = \frac{r_{\ell,t}^2}{ \sigma_{\ell,t}^2(\btheta)} + \log {\sigma}_{\ell,t}^2(\btheta),
$$
where ${\sigma}_{\ell,t}^2(\btheta)$ is defined in Section 2. Following the proof of Theorem 7.1 in Francq and Zokoian (2009), we shall establish the following results:
\begin{description}
\item[(a1)] $ \sup_{1 \le \ell \le p_n}\sup_{\bbtheta \in \bbLambda}| Q_{\ell,n}(\btheta) - \wt{Q}_{\ell,n}(\btheta)| \to  0, a.s.,$ as $n \to \infty$;
\item[(a2)] If there exists some $t$ such that $\wt\sigma_{\ell,t}^2(\btheta) = \wt\sigma_{\ell,t}^2(\btheta_{\ell,0})$ a.s. in $\bbP_{\btheta_{\ell,0}}$, then $\btheta = \btheta_{\ell,0}$;
\item[(a3)] $\bbE_{\bbtheta_{\ell,0}}|\wt v_{\ell,t}(\btheta_{\ell,0})| < \infty$, and if $\btheta \neq \btheta_{\ell,0}$, $\bbE_{\bbtheta_{\ell,0}}|\wt v_{\ell,t}(\btheta)| > \bbE_{\bbtheta_{\ell,0}}|\wt v_{\ell,t}(\btheta_{\ell,0})|$;
\item[(a4)] For any $\btheta \neq \btheta_{\ell,0}$, there exists a neighbourhood $U(\btheta)$ such that
$$
\lim \inf_{n \to \infty} \inf_{\bbtheta^* \in U(\bbtheta)} Q_{\ell,2}(\btheta) > \bbE_{\bbtheta_{\ell,0}} \wt v_{\ell,2}(\btheta_{\ell,0}), a.s.
$$
\end{description}

By the proof of Theorem 7.1 in Francq and Zokoian (2009), we only need to prove (a1).  Denote
\begin{eqnarray*}
\underline{\wt{\bsigma}}_{\ell,t}^2(\btheta)  = \left(\begin{array}{cc}
\wt \sigma_{\ell,t}^2(\btheta) \\
\wt \sigma_{\ell,t-1}^2(\btheta) \\
\vdots \\
\wt \sigma_{\ell,t - m+1}^2(\btheta)
\end{array}\right),
\underline{\wt{\bc}}_{\ell,t}(\btheta)  = \left(\begin{array}{cc}
\alpha_{0} + \sum_{j =1}^m\alpha_{j} \epsilon_{\ell,t-j}^2 \\
0 \\
\vdots \\
0
\end{array}\right),
\bB_{\ell}  = \left(\begin{array}{cccc}
\gamma_{1} &  \gamma_{2} & \cdots & \gamma_{s}\\
1 & 0 & \cdots & 0 \\
\vdots & \vdots & \vdots & \vdots \\
0 & \cdots & 1 & 0
\end{array}\right).
\end{eqnarray*}
We have the relationship $\underline{\wt{\bsigma}}_{\ell,t}^2 = \underline{\wt{\bc}}_{\ell,t} + \bB_{\ell}\underline{\wt{\bsigma}}_{\ell,t-1}^2.$
The condition (B2) and the compactness of $\bLambda$ implies that  $\rho = \sup_{\btheta \in \bLambda} \rho(\bB_{\ell}) < 1$, where $\rho(\bB)$ means the spectral radius of $\bB$.
Furthermore, $\underline{\wt{\bsigma}}_{\ell,t}^2$ can be expressed as
$$
\underline{\wt{\bsigma}}_{\ell,t}^2 = \sum_{k = 0}^{t-1}\bB^{k}_{\ell} \underline{\wt{\bc}}_{\ell,t-k} + \bB_{\ell}^t \underline{\wt{\bsigma}}_{\ell,0}^2.
$$
Let ${\underline{\bsigma}}_{\ell,t}^2(\btheta)$ be the vector obtained by replacing $\wt{\sigma}_{\ell,t-i}^2(\btheta)$ by ${{\sigma}}_{\ell,t-i}^2(\btheta)$ in $\underline{\wt{\bsigma}}_{\ell,t}^2(\btheta)$, and let ${\underline{\bc}}_{\ell,t}$ be the vector obtained by replacing $\epsilon_{\ell,t-i}^2$ by
$r_{\ell,t-i}^2$ and $r_{\ell,1}^2 ,\cdots, r_{\ell,2-m}^2$ by the initial values. Then we have
$$
{\underline{\bsigma}}_{\ell,t}^2 = \sum_{k = 0}^{t-1}\bB^{k}_{\ell} {\underline{\bc}}_{\ell,t-k} + \bB_{\ell}^t {\underline{\bsigma}}_{\ell,0}^2.
$$
Denote $\tilde{d}_{\ell} = \sup_{t \le n}|r_{\ell,t} - \epsilon_{\ell,t}|$. Then, if $t \ge m+1$,
\begin{eqnarray*}
 \|\widetilde{\underline{\bc}}_{\ell,t} - \underline{\bc}_{\ell,t}\| \le \left | \sum_{j = 1}^m \alpha_{j}(r^2_{\ell,t-j} - \epsilon_{\ell,t-j}^2) \right | \le \tilde{d}^2_{\ell} + 2\tilde{d}_{\ell}
\sum_{j = 1}^m \alpha_{j}\left |\epsilon_{\ell,t-j}\right |.
\end{eqnarray*}
As a result, for $t \ge m+1$, we obtain that
\begin{eqnarray*}
\|\widetilde{\underline{\bsigma}}_{\ell,t}^2 - \underline{\bsigma}_{\ell,t}^2\|
&\le & \left\|\sum_{k = 0}^{t - m + 1}\bB^{k}_{\ell} (\widetilde{\underline{\bc}}_{\ell,t-k} - \underline{\bc}_{\ell,t-k}) \right \|+
\left\|\sum_{k = t - m + 2}^{t-1}\bB^{k}_{\ell}(\widetilde{\underline{\bc}}_{\ell,t-k} - \underline{\bc}_{\ell,t-k}) \right \| \\
& & + \left\|\bB_{\ell}^t (\widetilde{\underline{\bsigma}}_{\ell,0}^2 -
\underline{\bsigma}_{\ell,0}^2) \right \|\\
&\le &  C \cdot \left( \tilde{d}^2_{\ell}  + \tilde{d}_{\ell} \sum_{k=0}^{t-1} \rho^k \sum_{j = 1}^m \alpha_{j}\left |\epsilon_{\ell,t-k-j}\right |+ \rho^t \|\widetilde{\underline{\bsigma}}_{\ell,0}^2 -
\underline{\bsigma}_{\ell,0}^2)\| \right ),
\end{eqnarray*}
for some constant $C>0$.
We thus have
\begin{eqnarray*}
\sup_{\btheta \in \bLambda}| Q_{\ell,n}(\btheta) - \widetilde{Q}_{\ell,n}(\btheta)|
&\le & n^{-1} \sum_{t = 2}^n \sup_{\btheta \in \bLambda} \left\{ \left |\frac{\tilde{\sigma}^2_{\ell,t} - \sigma^2_{\ell,t}}{\tilde{\sigma}^2_{\ell,t} \sigma^2_{\ell,t}} \right | \epsilon^2_{\ell,t} + \left|\log \frac{\sigma_{\ell,t}^2}{\tilde{\sigma}_{\ell,t}^2}\right|\right\} \\
&\le & \frac{1}{\alpha_{L}^2} \cdot C \cdot \left (\tilde{d}^2_{\ell}  + \tilde{d}_{\ell} + {n^{-1}}\sum_{t=2}^n \rho^t \epsilon_{\ell,t}^2 \right ) + \frac{1}{\alpha_{L}} \cdot C \cdot {n^{-1}}\sum_{t=2}^n \rho^t,
\end{eqnarray*}
where $\alpha_L = \inf_{\btheta \in \bLambda} |\alpha_{\ell,0}|$.  Note that $ \tilde{d}_{\ell} \le  C \cdot (h_1^2 + \delta_{3n}), a.s.$ and $\sup_{\ell \le p_n}\bbE \epsilon_{\ell,t}^{2d} < \infty$ implies that $\rho^t \epsilon_{\ell,t}^2 \to 0, a.s.$ Then
$\sup_{1\le \ell \le p_n}\sup_{\btheta \in \bLambda}| Q_{\ell,n}(\btheta) - \widetilde{Q}_{\ell,n}(\btheta)| \to 0, a.s.$, and part (a) follows.

Next, we consider the convergence rate of $\sup_{1 \le \ell \le p_n}\|\wh{\btheta}_{\ell} - \btheta_{\ell,0}\|$. The proof of this part is based on a standard Taylor expansion of $\widetilde{Q}_{\ell,n}(\btheta)$ at $\btheta_{\ell,0}$. Since $\wh{\btheta}_{\ell}$ converges to $\btheta_{\ell,0}$, which lies in the interior of the parameter space, we thus have
\begin{eqnarray*}
0 &= & n^{-1}\sum_{t = 2}^n \frac{\partial{v}_{\ell,t}(\wh{\btheta}_{\ell})}{\partial \btheta}  \\
  &= & n^{-1}\sum_{t = 2}^n \frac{\partial {v}_{\ell,t}(\btheta_{\ell,0})}{\partial \btheta}  +
 \left(\frac{1}{n}\sum_{t=2}^n \frac{\partial^2 {v}_{\ell,t}(\btheta^*_{\ell})}{\partial \btheta \partial \btheta^{\T}} \right)
 \cdot  (\wh{\btheta}_{\ell} - \btheta_{\ell,0}),
\end{eqnarray*}
where $\btheta_{\ell}^*$ is between $\wh{\btheta}_{\ell}$ and $\btheta_{\ell,0}$.
Suppose we have shown that there exist two positive constants $C_1$ and $C_2$ such that
\begin{eqnarray}
\label{lemma2-1}
\bbP \left \{ \sup_{1 \le \ell \le p_n} \left  \| \frac{1}{n}\sum_{t = 2}^n \frac{\partial {v}_{\ell,t}(\btheta_{\ell,0})}{\partial \btheta} \right \| > C_1 (h_1^2 + \delta_{3n}) \right \} = O\left( \frac{1}{n^{1 + \varepsilon}}\right),
\end{eqnarray}
and
\begin{eqnarray}
\label{lemma2-2}
 \bbP \left\{ \inf_{1 \le \ell \le p_n} \inf_{\btheta \in V(\btheta_0)}  \lambda_{\min}
\left ( \sum_{t=2}^n \frac{\partial^2 {v}_{\ell,t}(\btheta)}{\partial \btheta \partial \btheta^{\T}} \right )\le n C_2 \right \} = O\left( \frac{1}{n^{1 + \varepsilon}}\right).
\end{eqnarray}
Denote
\begin{eqnarray*}
\mathcal{A}_n = \left \{ \inf_{1 \le \ell \le p_n} \inf_{\btheta \in V(\btheta_{\ell,0})}\lambda_{\min} \left ( n^{-1}\sum_{t=2}^n \frac{\partial^2 {v}_{\ell,t}(\btheta )}{\partial \btheta \partial \btheta^{\T}} \right) > C_2 \right \},
\end{eqnarray*}
where $C_2$ is defined in (\ref{lemma2-2}). Then, for each $x > 0$, 
\begin{eqnarray}
\bbP \left \{ \sup_{1 \le \ell \le  p_n}\left\|\wh{\btheta}_{\ell} - \btheta_{\ell,0} \right \| > x \right\} \le
 \bbP \left\{
\sup_{1 \le \ell \le p_n}\left \| \sum_{t = 2}^n \frac{\partial {v}_{\ell,t}(\btheta_{\ell,0})}{\partial \btheta } \right\| > n C_2 x \right \} + \bbP (   \mathcal{A}_n^C).
\end{eqnarray}
Take $x = C_1 (h_1^2 + \delta_{3n})/C_2$ and the proof of Theorem 1(IV) follows immediately from (\ref{lemma2-1}) and (\ref{lemma2-2}). 

Now we prove (\ref{lemma2-1}) and (\ref{lemma2-2}). To establish (\ref{lemma2-1}) and (\ref{lemma2-2}), it suffices to prove the following five parts:
\begin{description}
\item[(b1)] There exists a constant $C > 0$ such that
\begin{eqnarray*}
\bbP \left \{ \sup_{1 \le \ell \le p_n} \left  \|\sum_{t = 2}^n \frac{\partial \wt{v}_{\ell,t}(\btheta_{\ell,0})}{\partial \btheta} \right \| > C n \delta_{3n} \right \} = o(1),
\end{eqnarray*}
\item[(b2)] There exists a constant $C > 0$ such that
\begin{eqnarray*}
\bbP \left \{ \sup_{1 \le \ell \le p_n} \left  |\sum_{t = 2}^n \frac{\partial \wt{v}_{\ell,t}(\btheta_{\ell,0})}{\partial \btheta}- \sum_{t = 2}^n \frac{\partial \tilde{v}_{\ell,t}(\btheta_{\ell,0})}{\partial \btheta}   \right | > C n (h_1^2 + \delta_{3n}) \right \} = O\left( \frac{1}{n^{1 + \varepsilon}}\right),
\end{eqnarray*}
\item[(b3)] There exists a constant $C > 0$ such that
\begin{eqnarray*}
\bbP \left \{ \inf_{1 \le \ell \le p_n}  \lambda_{\min} \left(\sum_{t=2}^n\frac{\partial^2 \wt{v}_{\ell,t}(\btheta_{\ell,0})}{\partial \btheta \partial \btheta^{\T}} \right ) \le n C \right \} = O\left( \frac{1}{n^{1 + \varepsilon}}\right),
\end{eqnarray*}
\item[(b4)] For any $C>0$, we have
\begin{eqnarray*}
\bbP \left \{ \sup_{1 \le \ell \le p_n} \sup_{\btheta \in V(\btheta_0)} \left \| \sum_{t=2}^n \frac{\partial^2 v_{\ell,t}(\btheta)}{\partial \btheta \partial \btheta^{\T}} -
\sum_{t=2}^n \frac{\partial^2 \tilde{v}_{\ell,t}(\btheta)}{\partial \btheta \partial \btheta^{\T}}
\right \| > nC \right \} = O\left( \frac{1}{n^{1 + \varepsilon}}\right),
\end{eqnarray*}
\item[(b5)] For each $i,j,k= 1,\cdots, m+ s + 1$, there exists a constant $C > 0$ and very small constant $c>0$ such that
\begin{eqnarray*}
\bbP \left \{ \sup_{1 \le \ell \le p_n} \sup_{\btheta \in V(\btheta_0)} \left | n^{-1}\sum_{t=2}^n \frac{\partial^2 \wt{v}_{\ell,t}(\btheta)}{\partial \theta_i \partial \theta_j \partial \theta_k}
\right | \le C n^{c} \right \} = O\left( \frac{1}{n^{1 + \varepsilon}}\right).
\end{eqnarray*}
\end{description}
It is not hard to see that (\ref{lemma2-1}) can be proved from (b1) and (b2) and (\ref{lemma2-2}) follows from (b3)-(b5). We now prove them separately. 

{\bf (b1).} It is easy to show that
\begin{eqnarray*}
\frac{\partial \wt v_{\ell,t}(\btheta)}{\partial \btheta} = \left ( 1 - \frac{\epsilon_{\ell,t}^2} {\wt \sigma_{\ell,t}^2(\btheta)}\right) \left ( \frac{1}{\wt \sigma_{\ell,t}^2(\btheta)} \frac{\partial \wt \sigma_{\ell,t}^2(\btheta)}{\partial \btheta}\right )
\end{eqnarray*}
and
\begin{eqnarray*}
\bbE \left\| \frac{\partial \wt v_{\ell,t}(\btheta_{\ell,0})}{\partial \btheta}\right \|^{d} < \infty.
\end{eqnarray*}
Note that $\{\epsilon_{\ell,t}, t \le n \}$ are strictly stationary and $\alpha-$mixing with geometric rate. (Also see Lindner (2009).) It follows from Theorem 2 (ii) of Liu, Xiao and Wu (2013) that, there exist  positive constants $C_1$, $C_2$ and $C_3$ such that for all $x > 0$,
\begin{eqnarray*}
\bbP \left \{ \left \|\sum_{t =2}^n \frac{\partial \wt v_{\ell,t}(\btheta_{\ell,0})}{\partial \theta} \right \| > x\right\} \le \frac{C_1 n}{x^{d}} + C_2 \exp\left( - \frac{C_3 x^2}{n^{1/2}}\right).
\end{eqnarray*}
Hence, by taking $x = C\delta_{2n}$ for a large constant $C>0$, we obtain that
\begin{eqnarray*}
\bbP \left \{ \sup_{1 \le \ell \le p_n}\left \|\sum_{t =2}^n \frac{\partial \wt v_{\ell,t}(\btheta_{\ell,0})}{\partial \btheta} \right \| > C \delta_{2n}\right\} &\le & \frac{C_1 n^{1 - d/2} p_n }{C^{d} (\log (n) )^{d/2}} + C_2 p_n \exp\left( - C_3 C^2 \log (n)\right ) \\
&\le  & O\left( \frac{1}{n^{1 + \varepsilon}}\right).
\end{eqnarray*}

{\bf (b2).} Similar to (a1) in this proof, we have that
\begin{eqnarray*}
\sup_{\bbtheta \in \bbLambda} \left \| \frac{\partial \tilde{\sigma}_{\ell,t}^2(\btheta)}{\partial \btheta}  -
\frac{\partial \sigma_{\ell,t}^2(\btheta)}{\partial \btheta}\right \| \le C(\tilde{d}_{\ell}^2 + \tilde{d}_{\ell} \sum_{k=0}^{t-1}\rho^k \sum_{j=1}^m |\epsilon_{t - k - j}|+ \rho^t ).
\end{eqnarray*}
We also obtain that
\begin{eqnarray*}
\wt{\sigma}_{\ell,t}^2\left| \frac{1}{\sigma_{\ell,t}^2} - \frac{1}{\tilde{\sigma}_{\ell,t}^2}\right| \le C(\tilde{d}_{\ell}^2 + \tilde{d}_{\ell} + \rho^t ),
\frac{\wt \sigma_{\ell,t}^2}{{\sigma}_{\ell,t}^2} \le 1 + C(\tilde{d}_{\ell}^2 + \tilde{d}_{\ell} + \rho^t ).
\end{eqnarray*}
As a result, for $i = 1, \cdots, m+s + 1$, the $i$-th component of the difference $\left | \frac{\partial v_{\ell,t}(\btheta_{\ell,0})}{\partial \btheta_i} - \frac{\partial \tilde{v}_{\ell,t}(\btheta_{\ell,0})}{\partial \btheta_i} \right |$ is bounded above by
\begin{eqnarray*}
&&\left | \frac{\partial \wt{v}_{\ell,t}(\btheta_{\ell,0})}{\partial \btheta_i} - \frac{\partial v_{\ell,t}(\btheta_{\ell,0})}{\partial \btheta_i} \right |\\
&&\le \left| \left ( \frac{\epsilon_{\ell,t}^2}{{\sigma}_{\ell,t}^2} -
\frac{\epsilon_{\ell,t}^2}{\wt {\sigma}_{\ell,t}^2}\right) \left( \frac{1}{\wt{\sigma}_{\ell,t}^2} \frac{\partial \tilde{\sigma}_{\ell,t}^2}{\partial \theta_i} \right) + \left(1 - \frac{\epsilon_{\ell,t}^2}{{\sigma}_{\ell,t}^2}\right)
\left( \frac{1}{\wt \sigma_{\ell,t}^2} - \frac{1}{{\sigma}_{\ell,t}^2} \right )  \frac{\partial {\sigma}_{\ell,t}^2}{\partial \theta_i} \right.\\
&& + \left.\left( 1 - \frac{\epsilon_{\ell,t}^2}{{\sigma}_{\ell,t}^2}\right) \frac{1}{{\sigma}_{\ell,t}^2}\left(\frac{\partial {\sigma}_{\ell,t}^2}{\partial \theta_i}  -
\frac{\partial \wt{\sigma}_{\ell,t}^2}{\partial \theta_i} \right)\right | (\btheta_{\ell,0}) +
\frac{\tilde{d}_{\ell}^2 +\tilde{d}_{\ell} |\epsilon_{\ell,t}|}{\sigma^2_{\ell,t}}\left| \frac{1}{\sigma_{\ell,t}^2}\frac{\partial {\sigma}_{\ell,t}^2}{\partial \theta_i} (\btheta_{\ell,0})\right |\\
& & \le C(\tilde{d}_{\ell}^2 + \tilde{d}_{\ell} + \rho^t ) (1 + \eta_{\ell,t}^2) \left|1+ \frac{1}{\wt \sigma_{\ell,t}^2(\btheta_{\ell,0})}\frac{\partial \wt\sigma_{\ell,t}^2(\btheta_{\ell,0})}{\partial \theta_i}\right |.
\end{eqnarray*}
Then it follows that, for $i = 1,\cdots,m+s+1,$
\begin{eqnarray*}
\left  |\sum_{t = 2}^n \frac{\partial v_{\ell,t}(\btheta_{\ell,0})}{\partial \theta_i}- \sum_{t = 2}^n \frac{\partial \tilde{v}_{\ell,t}(\btheta_{\ell,0})}{\partial \theta_i} \right |
&\le& C(\tilde{d}_{\ell}^2 + \tilde{d}_{\ell}) \sum_{t=2}^n (1 + \eta_{\ell,t}^2) \left|1+ \frac{1}{\wt \sigma_{\ell,t}^2(\btheta_{\ell,0})}\frac{\partial \wt \sigma_{\ell,t}^2(\btheta_{\ell,0})}{\partial \theta_i}\right | \\
&& + C\sum_{t=2}^n\rho^t (1 + \eta_{\ell,t}^2) \left|1+ \frac{1}{\wt \sigma_{\ell,t}^2(\btheta_{\ell,0})}\frac{\partial \wt \sigma_{\ell,t}^2(\btheta_{\ell,0})}{\partial \theta_i}\right |.
\end{eqnarray*}
By Markov and bulkholder inequalities for martingales, we claim that there exists a constant $C> 0$ such that
\begin{eqnarray*}
\bbP \left\{\sup_{1 \le \ell \le p_n}\sum_{t=2}^n\rho^t (1 + \eta_{\ell,t}^2) \left|1+ \frac{1}{\wt \sigma_{\ell,t}^2(\btheta_{\ell,0})}\frac{\partial \wt \sigma_{\ell,t}^2(\btheta_{\ell,0})}{\partial \theta_i}\right | > C  n^{1/2}\right \} = O\left(\frac{1}{n^{1+ \varepsilon}}\right).
\end{eqnarray*}
and
\begin{eqnarray*}
\bbP \left\{\sup_{1 \le \ell \le p_n}\sum_{t=2}^n (1 + \eta_{\ell,t}^2) \left|1+ \frac{1}{\wt \sigma_{\ell,t}^2(\btheta_{\ell,0})}\frac{\partial \wt \sigma_{\ell,t}^2(\btheta_{\ell,0})}{\partial \theta_i}\right | > C n \right \} = O\left(\frac{1}{n^{1+ \varepsilon}}\right).
\end{eqnarray*}
Note that $\sup_{\ell \le p_n}|\tilde{d}_{\ell}| = \bar{O}_{a.s}(h_1^2 + \delta_{3n})$. Hence, it follows that there exists a constant $C>0$ such that
\begin{eqnarray*}
\bbP \left \{ \sup_{1 \le \ell \le p_n} n^{-1}\left \|\sum_{t = 2}^n \frac{\partial v_{\ell,t}(\btheta_{\ell,0})}{\partial \bbtheta}- \sum_{t = 2}^n \frac{\partial \tilde{v}_{\ell,t}(\btheta_{\ell,0})}{\partial \bbtheta} \right \| > C(h_1^2 + \delta_{3n}) \right \} = O\left(\frac{1}{n^{1+ \varepsilon}}\right),
\end{eqnarray*}
and part (b2) follows.

{\bf (b3).}
$n^{-1}\sum_{t=2}^n \frac{\partial^2 \wt v_{\ell,t}(\btheta_{\ell,0})}{\partial \btheta \partial \btheta^{\T}}
$ can be expressed as
\begin{eqnarray*}
n^{-1}\sum_{t=2}^n \frac{\partial^2 \wt v_{\ell,t}(\btheta_{\ell,0})}{\partial \btheta \partial \btheta^{\T}}
= n^{-1}\sum_{t=2}^n \left \{\frac{\partial^2 \wt v_{\ell,t}(\btheta_{\ell,0})}{\partial \btheta \partial \btheta^{\T}} -
\bbE \left(\frac{\partial^2 \wt v_{\ell,t}(\btheta_{\ell,0})}{\partial \btheta \partial \btheta^{\T}}\right )\right \} +
\bbE \left \{  \frac{\partial^2 \wt v_{\ell,t}(\btheta_{\ell,0})}{\partial \btheta \partial \btheta^{\T}} \right\}.
\end{eqnarray*}
Note that $\inf_{1 \le \ell \le p_n}\bbE \left \{  \frac{\partial^2 \wt v_{\ell,t}(\btheta_{\ell,0})}{\partial \btheta \partial \btheta^{\T}} \right\}$ is positive definite. It suffices to show that, for any constant $c>0$,
 \begin{eqnarray*}
\bbP \left \{ \sup_{1 \le \ell \le p_n} n^{-1}\left |\sum_{t=2}^n  \left \{\frac{\partial^2 \wt v_{\ell,t}(\btheta_{\ell,0})}{\partial \btheta \partial \btheta^{\T}} -
\bbE \left(\frac{\partial^2 \wt v_{\ell,t}(\btheta_{\ell,0})}{\partial \btheta \partial \btheta^{\T}}\right)\right \}\right| > c \right \} = O\left(\frac{1}{n^{1+ \varepsilon}}\right).
 \end{eqnarray*}
Similar to (b1), we claim that there exist three positive constants $C_1$, $C_2$ and $C_3$ such that
\begin{eqnarray*}
& & \bbP \left \{ \sup_{1 \le \ell \le p_n} n^{-1} \left |\sum_{t=2}^n  \left \{\frac{\partial^2 \wt v_{\ell,t}(\btheta_{\ell,0})}{\partial \btheta \partial \btheta^{\T}} -
\bbE \left (\frac{\partial^2 \wt v_{\ell,t}(\btheta_{\ell,0})}{\partial \btheta \partial \btheta^{\T}}\right )\right \} \right |> c \right \}\\
& &\le C_1\frac{p_n n}{(nc)^{d}} + C_2 p_n \exp(- C_3 n^2 c^2) = O\left(\frac{1}{n^{1+ \varepsilon}}\right).
 \end{eqnarray*}
Part (b3) follows.

 {\bf (b4)} and {\bf (b5)}. Together with the proof of (c) in Theorem 7.2 of Francq and Zakoian(2011), the proofs of these two parts can be proved in a similar fashion to (b2) and (b3).


 \begin{center}
 {\bf \large Appendix D: Proof of Theorem 2}
 \end{center}

Define $\wh{\bE}_n = \wh{\bPhi}(\bX_n^{\T}\wh{\bbeta}) - \bPhi(\bX_n^{\T} \bbeta)$,
$
\wh{\bF}_n = \wh{\bSigma}_{x}(\bX_n) - \bSigma_{x}(\bX_n).
$
The difference $\wh{\cov}(Y_{n+1}|\mathcal{F}_n) - \cov(Y_{n+1}|\mathcal{F}_n)$ can be decomposed into four parts:
\begin{eqnarray*}
\wh{\cov}(Y_{n+1}|\mathcal{F}_n) - \cov(Y_{n+1}|\mathcal{F}_n)
& = & \wh{\bE}_n \wh{\bSigma}_{x}(\bX_n) \wh{\bE}_n^{\T} 
+ \bPhi(\bX_n^{\T}\bbeta) \wh{\bF}_n \{\bPhi(\bX_n^{\T} \bbeta )\}^{\T} + \left (\wh{\bSigma}_{0, n} - \bSigma_{0, n}\right )\\
& & + \left (\bPhi(\bX_n^{\T}\bbeta)\wh{\bSigma}_{x}(\bX_n) \wh{\bE}_n^{\T} + \wh{\bE}_n \wh{\bSigma}_{x}(\bX_n) \left\{\bPhi(\bX_n^{\T}\bbeta) \right\}^{\T}\right )  .
\end{eqnarray*}
We thus bound $\left\|\wh{\cov}(Y_{n+1}|\mathcal{F}_n) - \cov(Y_{n+1}|\mathcal{F}_n)\right\|_{\bSigma}^2 
$ by
\begin{eqnarray*}
&&4 \left\| \wh{\bE}_n \wh{\bSigma}_{x}(\bX_n ) \wh{\bE}_n^{\T}\right\|_{\bSigma}^2 + 
4 \left \|\bPhi(\bX_n^{\T}\bbeta) \wh{\bF}_n \{\bPhi ( \bX_n^{\T} \bbeta )\}^{\T} \right\|_{\bSigma}^2
+ 4 \left\|\wh{\bSigma}_{0, n} - \bSigma_{0, n}\right\|_{\bSigma}^2 \\
& &  +  4 \left\|\bPhi(\bX_n^{\T}\bbeta)\wh{\bSigma}_{x}(\bX_n) \wh{\bE}_n^{\T} + \wh{\bE}_n \wh{\bSigma}_{x}(\bX_n) \left\{\bPhi(\bX_n^{\T}\bbeta) \right\}^{\T}\right\|_{\bSigma}^2.
\end{eqnarray*}

To bound these terms, we first introduce the following two lemmas.

{\bf Lemma D.1.} Suppose that Assumptions (A1)-(A5), (B1)-(B4) and (C1)-(C4) in Appendix A hold. Then there exists a large $C > 0$ such that
\begin{description}
\item (i)
$$
P\left \{ \left\|\wh{\bE}_n \right \|_F^2 > C p_n\left(h_1^4 + \frac{\log (n)} {nh_1} \right)\right\}
\le O\left({1\over n^{1 + \varepsilon}}\right).
$$
\item (ii)
$$
P\left \{\left \|\wh{\bF}_n \right \|_F^2 > C \left(h_2^4 + \frac{\log (n)} {nh_2^q} \right)\right \}
\le O\left({1\over n^2}\right).
$$
\end{description}

{\bf Proof of Lemma D.1.} (i) Observe that
 \begin{eqnarray*}
\wh{\bPhi}(\bX_n^{\T}\wh{\bbeta})  - \bPhi(\bX_n^{\T}\bbeta) =  \bPhi'(\bX_n^{\T}\bbeta^* )  \bX_n^{\T} (\wh{\bbeta} - \bbeta)
+ \left(\wh{\bPhi}(\bX_n^{\T}\wh{\bbeta} )  - \bPhi (\bX_n^{\T}\wh{\bbeta}) \right),
\end{eqnarray*}
where $\bbeta^*$ is between $\wh{\bbeta}$ and $\bbeta$. As a result,
$$
\|\wh{\bE}_n\|_F^2 \le  2\|\sup_{z \in \mathcal{Z}}\bPhi'(z)\|_2^2\cdot \|\bX_n\|^2 \cdot \|\wh{\bbeta} - \bbeta\|_2^2 + 2\cdot \sup_{z \in \mathcal{Z}} \left\|\wh{\bPhi}(z)  - \bPhi(z)\right\|^2_F.
$$
Note that $\|\sup_{z \in \mathcal{Z}}\bPhi'(z)\|_F^2 \cdot \|\bX_n\|_2= O(p_n).$ Therefore, part (i) follows from Theorem 1(I) and (III).

(ii) Let $\tK_{h_2,t}(\bu) = \tK_{h_2}(\bX_{t-1} - \bu)$ and  $\varphi(\bX_t)$ be a bounded function uniformly over $\bX_t \in \mathcal{X}$. By following the proof of Theorem 5.3 in Fan and Yao (2003), we can see that there exists a large $C> 0$ such that
\begin{eqnarray*}
\bbP\left\{\sup_{\bu \in \mathcal{X}} \frac{1}{n}\left| \sum_{t=2}^n \varphi(\bX_{t})\tK_{h_2,t}(\bu) - \bbE \left\{\varphi(\bX_t) \tK_{h_2,t}(\bu)\right\} \right| > C \sqrt{\log ( n) \over n h_2^q}\right \}
\le O\left({1\over n^2}\right).
\end{eqnarray*}
By setting $\varphi(\bX_t) = 1,X_j,X_j X_k,(j, k =1,\cdots,q)$, part (ii) follows.

{\bf Lemma D.2.} Suppose that Assumptions (A1)-(A5), (B1)-(B4) and (C1) and (C3) in Appendix A hold. Then there exists $C > 0$ and small $\varepsilon >0$ such that
\begin{eqnarray*}
\bbP\left \{\sup_{1\le \ell \le p_n}\left|\frac{\wh{\sigma}^2_{\ell,n+1} - \sigma^2_{\ell,n+1}}{\sigma_{\ell,n+1}^2}\right| > C \left ( h_1^2 +  \delta_{3n}\right )\right \}
\le O\left({1\over n^{1 + \varepsilon}}\right ).
\end{eqnarray*}

{\bf Proof of Lemma D.2.} Let $\bB(i,j)$ be the $(i,j)$th element of the matrix $\bB$ and $\bA(i)$ be the $i$th entry of a vector $\bA$. The conditional covariance $\wh{\sigma}^2_{\ell,n+1}$ can be expressed as
\begin{eqnarray*}
\wh{\sigma}^2_{\ell,n+1} = \sum_{k = 0}^{n}\wh{\bB}^{k}_{\ell}(1,1) \wh{\underline{\bc}}_{\ell,n+1-k}(1) + \sum_{i=1}^{s}\wh{\bB}_{\ell}^{n+1}(1,i) \wh{\underline{\bsigma}}_{\ell,0}^2(i),
\end{eqnarray*}
where $\wh{\bB}_{\ell}$ is the matrix obtained by replacing $\wh{\gamma}_{\ell,j}$ by $\gamma_{\ell,j}$ in $\bB_{\ell}$ and $\wh{\underline{\bc}}_{\ell,t}$ and $\wh{\underline{\bsigma}}_{\ell,0}^2$ are defined accordingly.
Note that the true conditional variance 
\begin{eqnarray*}
\sigma^2_{\ell,n+1} = \sum_{k = 0}^{n}\bB^{k}_{\ell}(1,1) \underline{\bc}_{\ell,n+1-k}(1) + \sum_{i=1}^{s}\bB_{\ell}^{n+1}(1,i) \underline{\bsigma}_{\ell,0}^2(i).
\end{eqnarray*}
   We thus have that
\begin{eqnarray*}
\wh{\sigma}^2_{\ell,n+1}  - \sigma^2_{\ell,n+1}
&=& \sum_{k = 0}^{n}\wh{\bB}^{k}_{\ell}(1,1) \left(\wh{\underline{\bc}}_{\ell,n+1-k}(1) - \underline{\bc}_{\ell,n+1-k}(1)\right ) + \sum_{k = 1}^{n} \left( \wh{\bB}^{k}_{\ell} - \bB^k_{\ell}\right )(1,1) \underline{\bc}_{\ell,n+1-k}(1) \\
&+&  \sum_{i=1}^{s} \left ( \wh{\bB}_{\ell}^{n+1}(1,i)
\wh{\underline{\bsigma}}_{\ell,0}^2(i) - \bB_{\ell}^{n+1}(1,i)
\underline{\bsigma}_{\ell,0}^2(i) \right )
= U_{\ell,1} + U_{\ell,2} + U_{\ell,3}.
\end{eqnarray*}

(a) Consider the term $U_{\ell,1}$ and observe that
 $
\|\wh{\underline{\bc}}_{\ell,t} - \underline{\bc}_{\ell,t}\| \le |\wh{\alpha}_{\ell,0} - \alpha_{\ell,0}| + \tilde{d}^2_{\ell} + 2\tilde{d}_{\ell}
\sum_{j = 1}^m \left |\epsilon_{\ell,t-j}\right |.
$
Then, there exists a constant $C>0$ such that
$$
|U_{\ell,1} | \le C  \big(|\wh{\alpha}_{\ell,0} - \alpha_{\ell,0}| + \tilde{d}^2_{\ell} + \tilde{d}_{\ell}
\sum_{j = 1}^m \sum_{k=1}^n \rho^k \left |\epsilon_{\ell,t-k -j}\right |\big ).
$$
Since $\sum_{j = 1}^m \sum_{k=1}^n \rho^k \left |\epsilon_{\ell,t-k -j}\right |/\sigma_{\ell,n+1}^2 $ is bounded and $\sigma_{\ell,n+1}^2 \ge \alpha_{\ell,0 } >0$, this means that
$$
\left|\frac{U_{\ell,1} } {\sigma_{\ell,n+1}^2}\right| \le C \big(|\wh{\alpha}_{\ell,0} - \alpha_{\ell,0}| + \tilde{d}_{\ell} ).
$$
and consequently, there exists a large constant $C> 0$ such that
\begin{eqnarray*}
\bbP\left\{ \sup_{1 \le \ell \le p_n}\left|\frac{U_{\ell,1} } {\sigma_{\ell,n+1}^2}\right| > C(h_1^2 + \delta_{3n})\right\} = O\left(\frac{1}{n^{1+ \varepsilon}}\right).
\end{eqnarray*}

(b) Consider the term $U_{\ell,2}$. Denote $ \hat{\delta}_{\ell} = \sup_{1 \le i \le s}|\hat{\gamma}_{\ell,i}-\gamma_{\ell,i}|/\gamma_{\ell,i}$. By the definition of $\wh{\bB}_{\ell}$ and $\bB$, it is seen that
$$
\left|\frac {\wh{\bB}^{k}_{\ell}(1,1) - \bB^k_{\ell}(1,1)}{\bB^k_{\ell}(1,1)}\right| \le
\max \{|(1 - \hat{\delta}_{\ell})^k -1|,|(1 + \hat{\delta}_{\ell})^k - 1|\} \le 2 \hat{\delta}_{\ell} k (1+\hat{\delta}_{\ell})^{k-1},
$$
for small $\hat{\delta}_{\ell}$. Note that $\sigma_{\ell,n+1}^2  \ge \alpha_{\ell,0} + \bB^k_{\ell}(1,1) \underline{\bc}_{\ell,n+1 - k}(1)$ and the relation $x/(1+x) \le x^\delta$ for all $x\ge 0$ and $\delta \in (0,1)$. We have that
\begin{eqnarray*}
\left|\frac{U_{\ell,2}}{\sigma_{\ell,n+1}^2} \right|
&\le & \sum_{k = 1}^{n} \left |\frac{\left( \wh{\bB}^{k}_{\ell} - \bB^k_{\ell}\right )(1,1)}{\bB^k(1,1)} \right|\frac{\bB^k(1,1)\underline{\bc}_{\ell,n+1-k}(1)}{\alpha_{\ell,0} + \bB^k_{\ell}(1,1) \underline{\bc}_{\ell,n+1 - k}(1)} \\
& \le & 2 \hat{\delta}_{\ell} \sum_{k = 1}^{n} k(1 + \hat{\delta}_{\ell})^k \rho^{k\delta}\underline{\bc}_{\ell,n+1-k}^{\delta}(1).
\end{eqnarray*}
Hence, by choosing a suitable but small $\delta$, it follows from Theorem 1(IV) that there exists a large positive constant $C$ such that
\begin{eqnarray*}
\bbP \left\{\sup_{1 \le \ell \le p_n}\left|\frac{U_{\ell,2}}{\sigma_{\ell,n+1}^2} \right| > C (h_1^2 + \delta_{3n}) \right\}
\le \bbP \left\{\sup_{1 \le \ell \le p_n} \hat{\delta}_{\ell} > C (h_1^2 + \delta_{3n}) \right\}  = O\left(\frac{1}{n^{1+ \varepsilon}}\right).
\end{eqnarray*}

(c) It is easy to see that
$
\|U_{\ell,3}\|/\sigma_{\ell,n+1}^2$ is bounded. Lemma D.2 follows.

\vspace{3ex}
{\bf Proof of Theorem 2.}

(a). Now we bound $\left\| \wh{\bE}_n \wh{\bSigma}_{x}(\bX_n) \wh{\bE}_n^{\T}\right\|_{\bSigma}^2$.
Observe that
$$p_n\left\| \wh{\bE}_n \wh{\bSigma}_{x}(\bX_n) \wh{\bE}_n^{\T}\right\|_{\bSigma}^2
 = \lambda^2_{\max}\left(\cov(Y_{n+1}|\mathcal{F}_n)^{-1}\right)\lambda_{\max}^2\left(\wh{\bSigma}_{x}(\bX_n)\right)\left\| \wh{\bE}_n\right\|_F^4.
$$
Hence, it follows from Lemma D.1 that there exists $C > 0$ such that
\begin{eqnarray*}
\bbP \left \{ \left\| \wh{\bE}_n \wh{\bSigma}_{x}(\bX_n) \wh{\bE}_n^{\T}\right\|_{\bSigma}^2  >  Cp_n\left ( h_1^8 + \frac{\log^2 (n)}{(nh_1)^2} \right )\right \} = O\left(\frac{1}{n^{1+\varepsilon}}\right ).
\end{eqnarray*}

(b). We bound $ \left\|\bPhi(\bX_n^{\T}\bbeta) \wh{\bF}_n \{\bPhi(\bX_n^{\T} \bbeta )\}^{\T}\right\|_{\bSigma}^2$.
Note that
$
\|\bPhi(\bX_n^{\T} \bbeta)^{\T} \left(\cov(Y_{n+1}|\mathcal{F}_n)\right)^{-1} \bPhi(\bX_n^{\T} \bbeta)\| = O(1).
$
Hence, we have that
$
\left\|\bPhi(\bX_n^{\T}\bbeta) \wh{\bF}_n \{\bPhi(\bX_n^{\T} \bbeta )\}^{\T}\right\|_{\bSigma}^2
\le O(p_n^{-1}) \|\wh{\bF}_n\|^2_F,
$
and consequently, by Lemma D.1, there exists $C>0$ such that
\begin{eqnarray*}
\bbP \left \{ \left\|\bPhi(\bX_n^{\T}\bbeta) \wh{\bF}_n \{\bPhi(\bX_n^{\T} \bbeta )\}^{\T}\right\|_{\bSigma}^2  >  Cp_n^{-1}\left ( h_2^4 + \frac{\log (n)}{nh_2^q} \right )\right \} = O\left(\frac{1}{n^2}\right ).
\end{eqnarray*}

(c). We bound $\left\|\wh{\bSigma}_{0, n} - \bSigma_{0, n}\right\|_{\bSigma}^2$. Note that
\begin{eqnarray*}
\left\|\wh{\bSigma}_{0, n} - \bSigma_{0, n}\right\|_{\bSigma}^2
&\le & \left\|\cov(Y_{n+1}|\mathcal{F}_n)^{-1/2}\left ( \wh{\bSigma}_{0,n} - \bSigma_{0, n} \right ) \cov(Y_{n+1}|\mathcal{F}_n)^{-1/2}\right\|_2^2 \\
&\le & \sup_{1\le \ell \le p_n}\left | \frac{\wh{\sigma}_{\ell,n+1}^2 - \sigma_{\ell,n+1}^2}{\sigma_{\ell,n+1}^2}\right |^2.
\end{eqnarray*}
Hence we obtain from Lemma D.2 that there exists $C > 0$ such that
\begin{eqnarray*}
\bbP \left \{ \left\|\wh{\bSigma}_{0, n} - \bSigma_{0, n}\right\|_{\bSigma}^2 >  C\left ( h_1^4 + \frac{\log (n)}{nh_1} \right )\right \} = O\left(\frac{1}{n^{1+\varepsilon}}\right ).
\end{eqnarray*}

(d). Now we bound $\left\|\bPhi(\bX_n^{\T}\bbeta)\wh{\bSigma}_{x}(\bX_n) \wh{\bE}_n^{\T} + \wh{\bE}_n\wh{\bSigma}_{x}(\bX_n) \{\bPhi(\bX_n^{\T}\bbeta) \}^{\T}\right\|_{\bSigma}^2$.
Note that for two $q \times q$ matrix $\bA$ and $\bB$, $\|\bA+\bB\|_F^2 \le 2(\|\bA\|_F^2 + \|\bB\|_F^2)$, $\|\bA\bB\|_F \le \|\bA\|_F\|\bB\|_F$ and $|\tr(\bA\bB)| \le \|\bA\|_F\|\bB\|_F$. We have that
\begin{eqnarray*}
&&p_n\left\|\bPhi(\bX_n^{\T}\bbeta)\wh{\bSigma}_{x}(\bX_n) \wh{\bE}_n^{\T} + \wh{\bE}_n\wh{\bSigma}_{x}(\bX_n) \{\bPhi(\bX_n^{\T}\bbeta) \}^{\T}\right\|_{\bSigma}^2 \\
& & \le  2 \|\cov(Y_{n+1}|\mathcal{F}_n)^{-1/2}\bPhi(\bx^{\T}\bbeta)\wh{\bSigma}_{\bx}(\bx) \wh{\bE}_n^{\T}\cov(Y_{n+1}|\mathcal{F}_n)^{-1/2}\|_F^2 \\
& & =  2 \tr \left(\wh{\bSigma}_{x}(\bX_n) \wh{\bE}_n^{\T}\cov(Y_{n+1}|\mathcal{F}_n)^{-1}\wh{\bE}_n\wh{\bSigma}_{x}(\bX_n ) \{\bPhi(\bX_n^{\T}\bbeta)\}^{\T}\cov(Y_{n+1}|\mathcal{F}_n)^{-1}\bPhi(\bX_n^{\T}\bbeta)\right) \\
&& \le 2q^2\|\wh{\bSigma}_{x}(\bX_n)\|_F^2 \lambda_{\max}\left(\cov(Y_{n+1}|\mathcal{F}_n)^{-1}\right) \lambda_{\max}\left(\{\bPhi(\bX_n^{\T}\bbeta)\}^{\T}\cov(Y_{n+1}|\mathcal{F}_n)^{-1}\bPhi(\bX_n^{\T}\bbeta) \right) \cdot \|\wh{\bE}_n\|_F^2.
\end{eqnarray*}
Hence, by Lemma D.1 , together with $\lambda_{\max}\big(\{\bPhi(\bX_n^{\T}\bbeta)\}^{\T}\cov(Y_{n+1}|\mathcal{F}_n)^{-1}\bPhi(\bX_n^{\T}\bbeta) \big )= O(1)$, it follows that  there exists $C>0$ such that
\begin{eqnarray*}
\bbP \left\{\left\|\bPhi(\bX_n^{\T}\bbeta)\wh{\bSigma}_{x}(\bX_n) \wh{\bE}_n^{\T} + \wh{\bE}_n\wh{\bSigma}_{x}(\bX_n) \{\bPhi(\bX_n^{\T}\bbeta) \}^{\T}\right\|_{\bSigma}^2> C
\left(h_1^4 + \frac{\log n}{nh_1}\right) \right\} \le O\left(\frac{1}{n^{1+ \varepsilon}} \right).
\end{eqnarray*}

Combining (a)-(d), Theorem 2 follows. This completes the proof of Theorem 2.

\section*{References}
\begin{singlespace}
\begin{description}

\item Bosq, D. (1996). Nonparametric statistics for stochastic processes (Vol. 110). New York: Springer.

\item Bickel, P. and Levina, E. (2008a). Covariance regularization by
thresholding.  {\it Ann. Statist.}, {\bf 36}, 2577-2604.

\item Bickel, P. and Levina, E. (2008b). Regularized estimation of large
covariance matrices.  {\it Ann. Statist.}, {\bf 36}, 199-227.

\item Carroll, R. J., Fan, J., Gijbels, I. and Wand, M.P. (1997).  Generalized
partially linear single-Index models.  {\it Journal of American Statistical
Association}, {\bf 92}, 477-489.

\item El Karoui, N. (2008). Operator norm consistent estimation of a large
dimensional sparse covariance matrices.  {\it Ann. Statist.}, {\bf 36},
2717-2756.

\item Fama, E. and French, K. (1992). 
The cross-section of expected stock returns.
{\it J. Finance} {\bf 47}, 427-465.

\item Fama, E. and French, K. (1993).
Common risk factors in the returns on stocks and bonds.
{\it J. Financ. Econom.} {\bf 33}, 3-56.

\item Fan, J., Fan, Y. and Lv, J. (2008). High dimensional covariance matrix
estimation using a factor model.  {\it J. Econometrics}, {\bf 147}, 186-197.

\item Fan, J., Liao, Y., and Mincheva, M. (2011). High dimensional covariance matrix estimation in approximate factor models. {\it Ann. Statist.}, {\bf 39(6)}, 3320 - 3356.

\item Fan, J. and Yao, Q. (2003). Nonlinear Time Series: Nonparametric and Parametric Methods. Springer.

\item Fan, J., Yao, Q. and Cai, Z. (2003).  Adaptive varying-coefficient linear
models. {\it Journal of Royal Statistical Society B}, {\bf 65}, 57-80.

\item Fan, J. and Zhang, W. (1999). Statistical estimation in varying
       coefficient models.  {\it Ann. Statist.}, {\bf 27},
       1491-1518.

\item Fan, J. and Zhang, W. (2000).  Simultaneous confidence bands and
hypothesis testing in varying-coefficient models.  {\it Scandinavian Journal
of Statistics}, {\bf 27}, 715-731.

\item Francq, C. and Zakoian, J. M. (2011). GARCH models: structure, statistical inference and financial applications. John Wiley \& Sons.

\item H\"{a}rdle, W., Hall, P. and Ichimura, H. (1993). Optimal smoothing in
single-index models. {\it Ann. Statist.}, {\bf 21}, 157-178.

\item Kong, E. and Xia, Y. (2014).  An adaptive composite quantile approach to
dimension reduction.  {\it Annals of Statistics}, {\bf 42}, 1657-1688.

\item Lam, C. and Fan J. (2009). Sparsistency and rates of convergence in large
covariance matrix estimation.  {\it Ann. Statist.}, {\bf 37}, 4254-4278.

\item Li, J. and Zhang, W. (2011). A semiparametric threshold model for
censored longitudinal data analysis.  {\it Journal of the American Statistical
Association}, {\bf 106}, 685-696.

\item Lindner, A. M. (2009). Stationarity, mixing, distributional properties and moments of $\GARCH(p,q)$ processes. In {\it Handbook of financial time series} (pp. 43-69). Springer Berlin Heidelberg.

\item Liu, W., Xiao, H. and Wu, W. B. (2013). Probability and moment inequalities under dependence. {\it Statist. Sinca.}, {\bf 23(3)}, 1257-1272.

\item Markowitz, H.M. (1952). Portfolio selection
{\it J. Finance}, {\bf 7}, 77-91.

\item Markowitz, H.M. (1959). Portfolio Selection: Efficient Diversification of Investments. 
John Wiley \& Sons, New Jersey.

\item Rothman, A., Levina, E. and Zhu, J. (2009). Generalized thresholding of
large covariance matrices. {\it J. Amer. Statist. Assoc.}, {\bf 104}, 177-186.

\item Sun, Y., Yan, H., Zhang, W. and Lu, Z. (2014). A semiparametric spatial
dynamic model.  {\it The Annals of Statistics}, {\bf 42}, 700-727.

\item Sun, Y., Zhang, W. and Tong, H. (2007). Estimation of the covariance
matrix of random effects in longitudinal studies.  {\it The Annals of
Statistics}, {\bf 35}, 2795-2814.

\item Wu, W. B. and Pourahmadi, M. (2003). Nonparametric estimation of large
covariance matrices of longitudinal data. {\it Biometrika}, {\bf 94}, 1a€¡°17.

\item Xia Y. and H\"{a}rdle, W. (2006).  Semi-parametric estimation of
partially linear single-index models.  {\it Journal of Multivariate Analysis},
{\bf 97}, 1162-1184.

\item Xia, Y. and Li, W.K. (1999). On single-index coefficient regression models.
{\it J. Amer. Statist. Assoc.}, {\bf 94}, 1275-1285.
\item Xia, Y., Tong, H., and Li, W. K. (2002). Single-index volatility models and estimation. {\it Statist. Sinica.}, {\bf 12(3)}, 785-799.

\item Yu, Y. and Ruppert, D. (2002).  Penalized spline estimation for partially
linear single-index models.  {\it Journal of the American Statistical
Association}, {\bf 97}, 1042-1054.

\item Zhang, W., Fan, J. and Sun, Y. (2009). A semiparametric model for cluster
data.  {\it The Annals of Statistics}, {\bf 37}, 2377-2408.

\end{description}
\end{singlespace}

\end{document}